\documentclass[pra,twocolumn,superscriptaddress]{revtex4-2}
\usepackage{empheq,amssymb,orcidlink,hyperref,upgreek}

\hypersetup{
        colorlinks=true,       % false: boxed links; true: colored links
        linkcolor=red,        % color of internal links
        citecolor=blue,     % color of links to bibliography
        urlcolor=blue}

\begin{document}
\title{Floquet engineering in hybrid magnetic quantum systems}
\author{Feng-Zhou Ji\orcidlink{0000-0003-4859-1535}}
\affiliation{School of Physics, Henan Normal University, Xinxiang 453007, China}
\author{Si-Yuan Bai\orcidlink{0000-0002-4768-6260}}
\affiliation{Key Laboratory of Quantum Theory and Applications of MoE, Lanzhou Center for Theoretical Physics, Key Laboratory of Theoretical Physics of Gansu Province, Gansu Provincial Research Center for Basic Disciplines of Quantum Physics, Lanzhou University, Lanzhou 730000, China}
\author{Wan-Li Yang}
\affiliation{State Key Laboratory of Magnetic Resonance and Atomic and Molecular Physics, Innovation Academy for Precision Measurement Science and Technology, Chinese Academy of Sciences, Wuhan 430071, China}
\author{Chun-Jie Yang\orcidlink{0000-0003-2137-6958}}
\email{yangchunjie@htu.edu.cn}
\affiliation{School of Physics, Henan Normal University, Xinxiang 453007, China}
\author{Jun-Hong An\orcidlink{0000-0002-3475-0729}}
\email{anjhong@lzu.edu.cn}
\affiliation{Key Laboratory of Quantum Theory and Applications of MoE, Lanzhou Center for Theoretical Physics, Key Laboratory of Theoretical Physics of Gansu Province, Gansu Provincial Research Center for Basic Disciplines of Quantum Physics, Lanzhou University, Lanzhou 730000, China}

\begin{abstract}
%The advancement of magnonics has facilitated the utilization of hybrid magnetic systems in quantum technologies. A hybrid magnetic lattice (HML) formed by an array of superconducting loops and magnetic particles, has been devised as a quantum bus to disseminate quantum resources among magnetic quantum entities (MQEs) serving as nodes of a quantum network. However, the HML also exerts a decoherence effect on the MQEs, which has the potential to impair its practical performance. By studying the non-Markovian dynamics of two MQEs comprised of either nitrogen-vacancy centers or magnon modes coupled to two independent HMLs, we propose a Floquet-engineering scheme by applying periodic driving on the MQEs to overcome the unwanted effect. It is revealed that the decoherence can be suppressed and a significant degree of entanglement can be maintained in the steady state, provided that a Floquet bound state exists within the quasienergy spectrum of the total system of each periodically driven MQE and its HML. This result enhances our ability to control hybrid magnetic systems and is beneficial for the application of HML in quantum networks.

The advancement of magnonics has facilitated the utilization of hybrid magnetic systems in quantum technologies. A hybrid magnetic lattice formed by an array of superconducting loops and magnetic particles has been devised as a quantum bus to disseminate quantum resources among magnetic quantum entities serving as nodes of a quantum network. However, the lattice also exerts a decoherence effect on the quantum entities, which impairs its practical performance. By studying the non-Markovian dynamics of nitrogen-vacancy centers and magnon modes coupled to two independent hybrid magnetic lattices, we propose a Floquet-engineering scheme via periodic driving on the quantum entities to suppress decoherence. We find that significant steady-state entanglement is preserved when a Floquet bound state exists in the quasienergy spectrum of the system consisting of each driven quantum entity and its lattice. This result enables a precise manipulation of hybrid magnetic systems and benefits their applications in quantum networks.
\end{abstract}
\maketitle
%%%%%%%%%%%%%%%%%%%%%%%%%%%%%%%%%%%%%%%%%%%%%%%%%%%%%

\section*{Introduction} \label{sec:level1}
Recent developments in the field of hybrid magnetic quantum systems have been marked by a significant surge in magnon-based quantum applications \cite{10.1016/j.physrep.2022.03.002,10.1038/nphys3347}. The compatibility of magnetic structures has been demonstrated to be excellent in ensuring that the internal magnetic quasi-particle modes arising from the collective excitations of high-density spins are seamlessly integrated with nanomechanical phonons \cite{PhysRevX.12.011060,PhysRevApplied.13.064001,PhysRevB.99.184442}, photons \cite{PhysRevB.101.014419,doi.org/10.1038/npjqi.2015.14,PhysRevB.105.094422,PhysRevLett.124.053602}, and transmon qubits \cite{PhysRevB.108.224416,PhysRevA.109.022442,PhysRevB.108.094430}. This integration is facilitated through various coupling mechanisms, including magnetostrictive effect \cite{PhysRevA.110.023507,Li_2019,PhysRevX.11.031053}, magneto-optic coupling \cite{PhysRevLett.127.087203,PhysRevApplied.19.014002,PhysRevA.104.023711,PhysRevLett.123.127202}, dipole-dipole interaction \cite{PhysRevB.105.075410,PhysRevResearch.4.043180}, and exchange interaction \cite{PhysRevA.103.052411,PhysRevA.103.063702}.
Those hybrid magnetic quantum platforms are suitable for diverse quantum tasks \cite{doi.org/10.1016/j.physrep.2022.06.001,10.1063/5.0020277,PhysRevLett.130.073602}, such as quantum interconnect \cite{10.1038/nature07127,PRXQuantum.2.040344,PhysRevA.108.043703,10.1088/1402-4896/ad0d8d,PRXQuantum.2.040344}, quantum transduction \cite{PhysRevB.93.174427,PhysRevLett.117.133602}, and quantum magnetometry \cite{10.1063/5.0024369,PhysRevApplied.16.034036}. Those extensive applications offer opportunities for the transformative development in the field of magnetic quantum information and quantum computing \cite{doi.org/10.1063/5.0157520,10.1088/1361-648X/abec1a,doi.org/10.1038/nphys3410}.

\begin{figure}
 \centering
\includegraphics[width=\columnwidth]{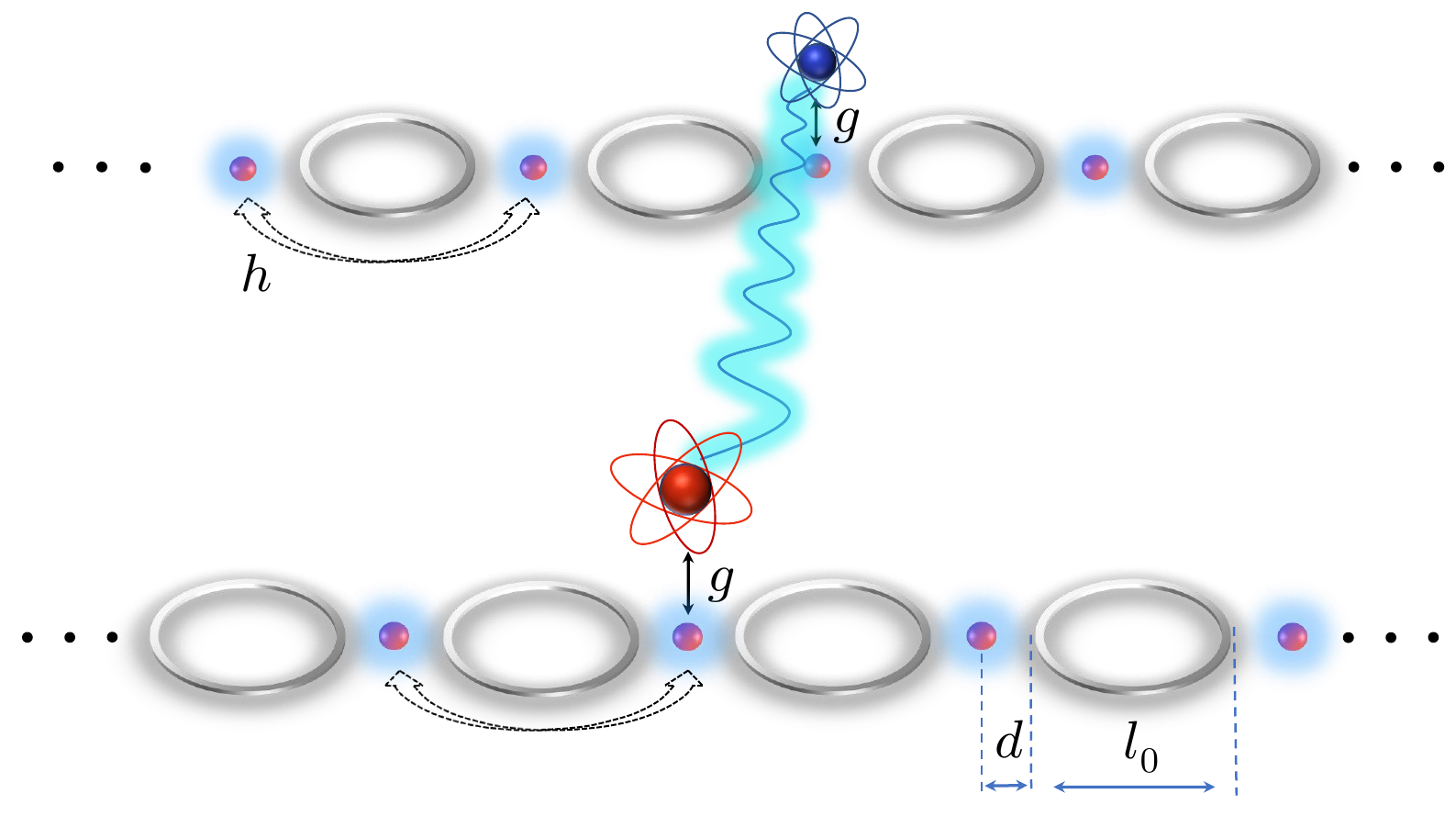}\\
  \caption{\textbf{Scheme of the system.} The hybrid magnetic lattices (HMLs) are both composed of a periodic arrangement of magnetic particles and superconducting loops separated in a distance of $d$. $l_0$ is the diameter of the loop. $h$ is the loop-mediated interaction between the nearest-neighbor magnetic particles. Two magnetic quantum entities (MQEs) denoted by color stars are independently coupled to two sets of HMLs with strength $g$. The MQE may be either a discrete- or a continuous-variable system periodically driven by external magnetic field. }\label{Fig1}
\end{figure}

The hybrid magnetic lattice (HML) \cite{PhysRevA.100.022343} composed of superconducting loops and magnetic particles has been proposed to mediate interactions between magnetic quantum entities (MQEs) over a long distance. Combined with the tunability of flux-mediated interactions supplied by superconducting loops \cite{PhysRevLett.129.037205}, the HML might facilitate plenty of applications in quantum information, including quantum sensing \cite{PhysRevLett.125.117701}, quantum gate \cite{PhysRevB.110.104416} and nonclassical magnetic quantum state generation \cite{PhysRevA.110.013711,PhysRevA.110.053710}. However, acting as a multimode environment, the HML itself causes an additional loss to the quantum system, which leads to the complete decay of quantum coherence in a long time \cite{PhysRevLett.125.247702,PRXQuantum.2.040314,PhysRevResearch.4.023221}. This loss introduces unwanted decoherence to the MQEs. In terms of practical applications \cite{doi.org/10.1016/j.jmmm.2020.166711,10.1088/1361-648X/ad6828}, persistent quantum correlation between the MQEs is desired, which requires an effective tool to suppress the decoherence. A widely used dynamical decoupling method to suppress the decoherence highly depends on precise control of the applied pulse sequences, which raises a high experimental requirement \cite{PhysRevLett.82.2417, PhysRevA.58.2733}. On the other hand, Floquet engineering, which exploits the external always-on periodic driving to control systems, has emerged as a versatile tool in modern quantum technologies \cite{PhysRevLett.117.250401,10.1080/00018732.2015.1055918,10.1080/23746149.2020.1870559}, e.g., in generating time crystal with long coherence time \cite{PhysRevLett.117.090402,PhysRevLett.126.163902,PhysRevB.109.174310} and discovering anomalous quantum phase in quantum many-body systems \cite{10.1088/1367-2630/17/9/093039,10.1038/s41567-019-0698-y}. Therefore, we expect to explore the constructive role of Floquet engineering in suppressing the decoherence of MQEs caused by the HML, which is helpful for the practical applications of HML in quantum technologies.

In the paper, we propose a periodic driving scheme to suppress the decoherence and maintain the quantum entanglement of bipartite MQEs independently coupled to two isolated HMLs. We find that the decoherence dynamics of the MQEs caused by the HMLs is determined by the features of the Floquet quasienergy spectrum of the total system formed by each periodically driven MQE and its local HML. With the formation of a Floquet bound state (FBS) in the quasienergy spectrum, a persistent quantum entanglement in tiny-amplitude oscillations synchronizing with the periodic driving between the bipartite MQEs is preserved in the long-time limit. Our results indicate that the Floquet engineering can rescue the bipartite entanglement from decoherence induced by the local HML environment. Our findings support the applications of HMLs in the design of magnetic quantum devices.

%The paper is organized as follows. In Sec. \ref{model}, we show the model of two periodically driven MQEs coupled to two independent HMLs. In Sec. \ref{dynamics}, we study the non-Markovian dynamics of the MQEs comprised of either nitrogen-vacancy (NV) centers or magnon modes. The dominant role of FBS in suppressing the decoherence and maintaining persistent entanglement between the MQEs is demonstrated. In Sec. \ref{entanglement preservation}, we provide numerical calculations on the dynamics, which verifies our analytical results. In Sec. \ref{conclusion}, we discuss and summarize our results.

\section*{Results}
\subsection*{System}\label{model} We consider two sets of subsystems, each of which consists of a periodically driven MQE coupled to a one-dimensional HML, as shown in Fig. \ref{Fig1}. The Hamiltonian of the whole system is
$\hat{H}(t)=\sum_{l=1}^{2}\hat{H}_{l}(t)$, with
\begin{equation}\label{mmm}
\hat{H}_l(t)=\hat{H}^{\text{sys}}_{l}(t)+\hat{H}^{\text{hml}}_{l}+\hat{H}^{\text{int}}_{l}.
\end{equation}
Here, $\hat{H}^{\text{sys}}_{l}$, $\hat{H}^{\text{hml}}_{l}$, and $\hat{H}^{\text{int}}_{l}$ are the Hamiltonian of the MQE, the HML, and their interaction in the $l$th subsystem, respectively. The HML is implemented by a periodic arrangement of superconducting loops and magnetic particles. The superconducting loop with diameter $l_0$ behaves as a single-mode resonator \cite{PhysRevApplied.5.044021}. The magnetic particle may be a yttrium iron garnet (YIG) sphere. Applying a magnetic field $\mathbf{B}_{z}=-B_{z}\mathbf{e}_{z}$ parallel to the plane of the loop, the magnetic particles with radius $R$ are uniformly magnetized with a magnetic dipole moment $\pmb{\mu}$. Each magnetized particle exerts an external flux on its superconducting loop. It leads to an effective interaction between the superconducting loops and the magnetic particles. After tracing out the electromagnetic mode of the loops, the loop-mediated indirect interactions among the magnon modes in the magnetic particles are generated. For large $B_{z}$ and under the condition of identical loops and magnetic particles, the Hamiltonian of the HML reads \cite{PhysRevA.100.022343}
\begin{equation}\label{hml}
\hat{H}_l^{\text{hml}}=\hbar\omega_c\sum_{n=1}^{N}\hat{f}^{\dagger l}_{n}\hat{f}_{n}^l-\hbar h\sum_{n=1}^{N-1}(\hat{f}^{\dagger l}_{n}\hat{f}_{n+1}^l+\text{H.c.}).
\end{equation}
The first term is the magnon modes in the magnetic particles with the annihilation operator $\hat{f}_{n}^l$ satisfying $[\hat{f}_{m}^l,\hat{f}^{\dagger l}_n]=\delta_{mn}$ and a common frequency $\omega_c$. The second term denotes the loop-mediated coupling between the magnons in the nearest-neighbor magnetic particles in a coupling strength $h$. The magnetic dipole-dipole coupling between the magnetic particles has been neglected because it is much smaller than the loop-mediated coupling when the size of the loop is large enough compared to the distance between it and the magnetic particles, i.e., $l_0\gg d$ in Fig. \ref{Fig1}. The Hamiltonian of Eq. \eqref{hml} describes the tunneling of the magnons as magnetic quasi-particles between the nearest YIG spheres mediated by a superconducting loop, which resembles the propagation of a photon in a photonic lattice \cite{PhysRevResearch.3.L022025,PhysRevA.97.023808}. In comparison with the photonic crystal, the feature of this magnetic lattice is that its frequency band can be tuned freely. To see this, we rewrite Eq. \eqref{hml} in momentum space. Applying the Fourier transform $\hat{f}_{n}^l=\frac{1}{\sqrt{N}}\sum_k \hat{\tilde f}^l_k \mathrm{e}^{\mathrm{i}kn}$ under the periodic boundary condition, we have $\hat{H}^{\text{hml}}_l=\sum_k\hbar\omega_{k}\hat{\tilde f}^{\dagger l}_k\hat{\tilde f}_{k}^l$, with the dispersion relation $\omega_{k}=\omega_{c}-2h\cos k$. In the large-$N$ limit, a continuous frequency band with width $4h$ centered at $\omega_c$ is formed in the region $\omega_{k}\in[\omega_{c}-2h,\omega_{c}+2h]$. Both the center frequency $\omega_c$ and the band width $4h$ can be controlled by the external magnetic field $B_z$ \cite{PhysRevA.100.022343}. Thus, a tunable band structure of the artificial one-dimensional magnetic crystal is obtained.

Another key feature of the magnetic quasi-particle based HML is its ability to couple to various MQEs, including solid spin and magnetic quasi-particle, by gradient magnetic field \cite{PhysRevLett.121.123604}. We investigate each HML interacting with a MQE via
\begin{equation}\label{int}
\hat{H}^{\text{int}}_l=-\hbar g(\hat{O}^{\dagger}_l\hat{f}_1^l+\text{H.c.}),
\end{equation}
where $\hat{O}$ is the annihilation operator of the MQE and $g$ is its coupling strength to the HML. Depending on its property, each MQE may be either a discrete-variable system or a continuous-variable one. The former can be either a magnetic dipolar emitter \cite{PhysRevLett.125.247702,PhysRevB.108.L180409} or a spin defect \cite{Yang_2019,PhysRevResearch.4.043180}, which interacts with the HML by magnetic dipolar interactions and can be modeled by a two-level system with $\hat{O}=\hat{\sigma}$. The latter can be a magnetic particle, which interacts with the HML through another superconducting loop and is described by another magnon mode with $\hat{O}=\hat{a}$. Equation \eqref{int} in the momentum space reads $\hat{H}^{\text{int}}_l=-\hbar \sum_k g_k(\hat{O}^{\dagger}_l\hat{\tilde f}_k^l+\text{H.c.})$, where $g_k=g \mathrm{e}^{\mathrm{i}k}/\sqrt{N}$.

This scheme opens up the possibility of controlling quantum entanglement between the MQEs and lays the foundation for realizing magnetic quantum network \cite{doi.org/10.1021/nl102066q,doi.org/10.1038/s41586-018-0200-5,10.1126/science.abg1919}, quantum simulation \cite{PhysRevA.107.053701,PhysRevB.87.174407}, and quantum information processing \cite{10.1063/5.0020277,doi.org/10.7567/1882-0786/ab248d}. Serving as an environment, the HML causes decoherence to the MQE, which is the main obstacle to exploring its applications. It has been found that the non-Markovian effect in strong-coupling regime plays a constructive role in suppressing decoherence \cite{PhysRevB.108.L180409}. However, the low saturation magnetization of the magnetic material and the difficulty in regulating the established structure of HML prevent the system from approaching the strong coupling \cite{PhysRevMaterials.3.034403}. Inspired by the progress that Floquet engineering by applying an external periodic driving on quantum systems has become a versatile tool of quantum control \cite{10.1080/23746149.2020.1870559,PhysRevA.102.060201,PhysRevA.91.052122,PhysRevLett.131.050801}, we explore the possibility of decoherence suppression by periodic driving. The Hamiltonian of the MQEs under the periodic driving is
\begin{equation}\label{sys}
\hat{H}^{\text{sys}}_l(t)=\hbar[\omega_0+A(t)]\hat{O}^{\dagger}_l\hat{O}_l,
\end{equation}
where $\omega_0$ is the eigen-frequency of the MQEs and $A(t)$ is the external driving. We adopt a piecewise periodic-driving protocol
\begin{equation}\label{step}
 A(t)=\left\{
\begin{array}{rcl}
\mathcal{F},   && jT \leq t < jT+t^{\prime}\\
0,   && jT+t^{\prime} \leq t < (j+1)T
\end{array},
\right.
\end{equation}
where $j$ are integer numbers, $\mathcal{F}$ and $T$ are the driving amplitude and period. Being widely used in generating nonequilibrium quantum states of matter, e.g., time crystal \cite{Zhang2017}, the protocol can be realized by applying a magnetic field to the MQEs. The choice of a two-segment periodic driving over a smooth one is motivated by the convenience of numerical calculation and carries no special implication for the theoretical validity.

\subsection*{Exact dynamics}\label{dynamics}
The non-Markovian dynamics of the MQEs is exactly solvable under the initial condition that the magnons of the two sets of the HMLs are in the vacuum state $|\{0_k\}_1,\{0_k\}_2\rangle$. When the MQEs are two-level systems, the reduced dynamics of such a discrete-variable system can be exactly derived by tracing over the degrees of freedom of HMLs from the unitary dynamics of the total system \cite{PhysRevB.108.L180409}. When the MQEs are magnons, the exact dynamics is obtained by the Feynman-Vernon's influence-functional theory in the coherent-state representation \cite{PhysRevA.76.042127}. Then, a general form of the exact master equation for both the discrete- and continuous-variable cases is (see Supplementary Note 1)
\begin{equation}
\dot{\rho}(t)=\sum_{l=1}^{2}\{-\mathrm{i}\Omega (t)[\hat{O}^{\dagger}_{l}\hat{O}_l,\rho(t)]+\Gamma (t)\check{\mathcal{L}}_{l}\rho (t)\},
\label{masterequation1}
\end{equation}
where $\check{\mathcal{L}}_{l}\cdot=2\hat{O}_{l}\cdot\hat{O}_{l}^{\dag}-\hat{O}_{l}^{\dag}\hat{O}_{l}\cdot-\cdot\hat{O}_{l}^{\dag}\hat{O}_{l}$ is the Lindblad superoperator, $\Omega(t)=-\text{Im}[\dot{c}(t)/c(t)]$ and $\Gamma(t)=-\text{Re}[\dot{c}(t)/c(t)]$ denote the renormalized frequency and decay rate of the quantum system.  The time-dependent parameter $c(t)$ is determined by
\begin{equation}\label{dynamics1}
  \dot{c}(t)+\mathrm{i}[\omega_0+A(t)]c(t)+\int_{0}^{t}d\tau F(t-\tau)c(\tau)=0
\end{equation}
under the initial condition $c(0)=1$, where $F(t-\tau)=\int d\omega J(\omega)\mathrm{e}^{-\mathrm{i}\omega (t-\tau)}$ is the correlation function and $J(\omega)=\sum_k|g_k|^2\delta(\omega-\omega_{k})$ is the spectral density of the HMLs. In the continuous limit of the HML frequencies, we have $\sum_k\rightarrow\int dk/(2\uppi N)$ and $F(t-\tau)=g^{2}\mathrm{e}^{-\mathrm{i}\omega_{c}(t-\tau)}\mathcal{J}_{0}(2h(t-\tau))$, where $\mathcal{J}_{0}$ is the zeroth-order Bessel function of the first kind. The convolution in Eq. \eqref{dynamics1} renders the dynamics non-Markovian. In the absence of the periodic driving and in the weak-coupling regime, we can obtain the Born-Markovian approximate solution of Eq. \eqref{dynamics1} as $c_{\text{MA}}(t)=\exp[-(\gamma+\mathrm{i}\Delta)t]$, with $\gamma=\uppi J(\omega_{0})$, $\Delta=\mathcal{P}\int_{0}^{\infty}\frac{J(\omega)}{\omega_{0}-\omega}d\omega$, and $\mathcal{P}$ being the Cauchy principal value. In the long-time limit, we have $c_{\text{MA}}(\infty)=0$. Thus, the MQEs experience an exponential decay to their ground state.

When the periodic driving is applied, the Born-Markovian approximation is generally invalid even in the weak-coupling regime. The solution of Eq. \eqref{dynamics1} is obtainable only via numerical calculations. On the other hand, Floquet theorem, as a useful tool to deal with the dynamics of periodically driven systems \cite{PhysRevLett.117.250401,10.1080/00018732.2015.1055918,10.1080/23746149.2020.1870559,PhysRevLett.117.090402,PhysRevLett.126.163902,PhysRevB.109.174310,10.1088/1367-2630/17/9/093039,10.1038/s41567-019-0698-y}, permits us to obtain the asymptotic solution of Eq. \eqref{dynamics1} in the long-time limit. According to Floquet theorem, the evolution operator of the total system $\hat{H}_l(t)$ consisting of the periodically driven MQE and its local HML is expanded as \cite{PhysRev.138.B979,PhysRevA.7.2203}
\begin{equation}
   \hat{\mathcal{U}}_l(t)=\sum_{n}\mathrm{e}^{-\mathrm{i}\epsilon_n t/\hbar}|u_n(t)\rangle\langle u_n(0)|,\label{evsp}
\end{equation}where $\epsilon_n$ called the Floquet quasi-energies are time-independent and $|u_n(t)\rangle$ called the quasi-stationary states satisfy $|u_n(t)\rangle=|u_n(t+T)\rangle$. They are determined by the Floquet equation
\begin{equation}
  [\hat{H}_l(t)-\mathrm{i}\hbar\partial_t]|u_n(t)\rangle=\epsilon_n|u_n(t)\rangle.  \label{fladt}
\end{equation}
It is easy to verify that $\mathrm{e}^{\mathrm{i}m\Omega t}|u_n(t)\rangle$ ($m$ being an arbitrary integer number and $\Omega=2\uppi/T$ being the driving frequency), which are the same states as $|u_n(t)\rangle$ up to a trivial phase factor, are also the eigenstates of Eq. \eqref{fladt}. However, their eigenvalues become $ \epsilon_n+m\hbar\Omega$. It means that the Floquet quasienergies are periodic with period $\hbar\Omega$. One generally chooses the first period, i.e., $[-0.5\hbar\Omega,+0.5\hbar\Omega]$ being called the first Brillouin zone (FBZ) of the quasienergies, to completely characterize the quansienergy spectrum. It can be proven that the solution of Eq. \eqref{dynamics1} is equivalent to (see Supplementary Note 2)
\begin{equation}
    c(t)=\langle \Phi(0)|\hat{\mathcal{U}}_l(t)|\Phi(0)\rangle,\label{cdss}
\end{equation}
where $|\Phi(0)\rangle\equiv\hat{O}_l^{\dag}|\varnothing_l,\{0_{k}\}_l\rangle$, with $|\varnothing\rangle$ being the ground state of the $l$th MQE. Therefore, $|c(t)|^2$ describes the initial-state fidelity of the evolved state from $|\Phi(0)\rangle$. Substituting Eq. \eqref{evsp} into Eq. \eqref{cdss} and using $|u_n(t)\rangle=|u_n(t+T)\rangle$, we have $c(t)=\sum_{n}\mathrm{e}^{-\mathrm{i}\epsilon_nt/\hbar}|\langle u_n(0)|\Phi(0)\rangle|^2$
in the stroboscopic dynamics at $t =mT$, with $m\in\mathbb{Z}$. It indicates that the dynamics of the MQEs governed by $c(t)$ is essentially determined by the feature of the quasienergy spectrum $\epsilon_n$ in the single-excitation subspace of the total system formed by each periodically driven MQE and its coupled HML. The quasienergy spectrum is generally composed of a continuous quasienergy band and a possibly formed discrete quasienergy $\epsilon^b$ residing in the band-gap regime. We call the eigenstate $|u^b(t)\rangle$ corresponding to the quasienergy $\epsilon^b$ FBS. In the long-time limit, the contribution from the continuous quasienergy band to $c(t)$ tends to zero due to the out-of-phase interference and only the component of the FBS survives. Therefore, we obtain the long-time solution of the dynamics at $t=mT$ as \cite{PhysRevLett.131.050801}
\begin{equation}
\lim_{t\rightarrow\infty}c(t)=\mathcal{Z}\mathrm{e}^{-\mathrm{i}\epsilon^{b}t/\hbar},\label{longtime2}
\end{equation}
where $\mathcal{Z}=|\langle u^{b}(0)|\Phi(0)\rangle|^2$. If the FBS is absent, then $\lim_{t\rightarrow\infty}c(t)=0$, which characterizes a complete decoherence. If the FBS is formed, then $c(t)$ tends to Eq. \eqref{longtime2}, which implies a decoherence suppression. Entirely absent in the Born-Markovian approximation, Eq. \eqref{longtime2} reveals that we can suppress the decoherence of the MQE caused by the HML via engineering the quasienergy spectrum of the total MQE-HML system to form the FBS. Efficiently overcoming the destructive impact of the HMLs on the MQEs, this result also enables us to distribute stable entanglement between spatially separated MQEs via the HMLs. This is helpful in designing functional quantum devices such as quantum networks \cite{10.1126/science.abg1919,PhysRevLett.127.183202,PRXQuantum.2.040344}.

To verify our result, we explicitly investigate the dynamics of two kinds of MQEs. We first consider that the MQEs are formed by two NV centers in diamonds as spin defects. The NV center consists of a substitutional nitrogen atom and an adjacent vacancy in the diamond. It has the ground triplet states $|m_s=0\rangle$ and $|m_s=\pm 1\rangle$, which have a zero-field splitting $D=2\uppi\times 2.87$ GHz. A static magnetic field $B_{\text{s}}$ is applied to remove the degenerate sublevels $|m_s=\pm 1\rangle$ with a Zeeman splitting $\delta =2g_e\mu_BB_\text{s}$, where $g_e\simeq 2$ and $\mu_B=14$MHz/mT are the NV’s Land\'{e} factor and Bohr magneton, respectively. Then applying a dichromatic microwave field $B_x^{\pm}(t)=B_0\cos(\omega_{\pm}t)$ on the NV center, the Hamiltonian in the rotating frame with the microwave frequencies $\omega_{\pm}$ is \cite{Yang_2019,PhysRevApplied.10.024011,PhysRevLett.125.153602}
\begin{equation}
\hat{H}_{\text{NV}}=\sum_{j=\pm}[\Delta_j|j\rangle\langle j|+\Lambda/2(|0\rangle\langle j|+|j\rangle\langle 0|)],
\end{equation}
where $\Delta_{\pm}=D\pm\frac{\delta}{2}-\omega_{\pm}$ and $\Lambda=g_e\mu_B B_0/\sqrt{2}$. Choosing the frequencies $\omega_\pm$ appropriately such that $\Delta_{+}=\Delta_-\equiv\Delta$, we obtain the eigenbasis of $\hat{H}_{\text{NV}}$ as a dark state $|d\rangle=(|+1\rangle-|-1\rangle)/\sqrt{2}$ and two dressed states $|g\rangle=\sin\theta|b\rangle-\cos\theta|0\rangle$ and $|e\rangle=\cos\theta|b\rangle+\sin\theta|0\rangle$, where $|b\rangle=(|+1\rangle+|-1\rangle)/\sqrt{2}$ and $\tan(2\theta)=\sqrt{2}\Lambda/\Delta$. Their eigenfrequencies are $\omega_d=\Delta$ and $\omega_{e/g}=(\Delta\pm\sqrt{\Delta^2+2\Lambda^2})/2$. Then we can see that $(\omega_d-\omega_g)-(\omega_e-\omega_d)=\Delta$. When $\Delta > 4h$, we can ensure that the transition $\omega_e - \omega_d$ lies within the HML's energy band by choosing a suitable $\Lambda$, while the other transition $\omega_d - \omega_g$ falls far outside this band. Then we can treat the NV center as a spin-1/2 system with $|\uparrow\rangle\equiv|e\rangle$ and $|\downarrow\rangle\equiv|d\rangle$ with transition frequency $\omega_0=\omega_e-\omega_d$. The interaction Hamiltonian of the $l$th NV center coupled to the magnon mode in the first magnetic particle of the $l$th HML under the rotating-wave approximation can be written
\begin{equation}\label{int}
\hat{H}^\text{int}_l=-\hbar g(\hat{\sigma}^{\dagger}_l\hat{f}_1^l+\text{H.c.}),
\end{equation}where $\hat{\sigma}^\dag=|\uparrow\rangle \langle \downarrow|$. Then we can modulate $\Lambda$ via tuning $B_0$ to realize the piecewise periodic-driving protocol. Being widely viewed as a two-level system, the NV center has a wide prospect in realizing quantum sensing and quantum computation. In the following, we show the constructive role of Floquet engineering in suppressing the decoherence in the NV centers and preserving the entanglement between two NV centers coupling to two independent HMLs. Solving the master equation \eqref{masterequation1} with $\hat{O}=\hat{\sigma}$ under the initial condition $\rho=|\Psi(0)\rangle\langle\Psi(0)|$ with $|\Psi(0)\rangle= (|\uparrow\uparrow\rangle+|\downarrow\downarrow\rangle)/\sqrt{2}$, we obtain the reduced density matrix of the two NV centers (see Supplementary Note 1)
\begin{eqnarray}\label{concurrence}
\rho (t) &=&\{[P_{t}|\uparrow\rangle \langle \uparrow|+(1-P_{t})|\downarrow\rangle \langle
\downarrow|]^{\otimes 2}+|\downarrow\rangle \langle \downarrow|^{\otimes 2}  \notag \\
&&+[c^{2}(t)|\uparrow\rangle \langle \downarrow|^{\otimes 2}+\text{H.c.}]\}/2,
\end{eqnarray}
where $P_{t}=|c(t)|^{2}$. In the absence of the FBS, $c(\infty)=0$ and thus $\rho(\infty)=|\downarrow\downarrow\rangle \langle\downarrow\downarrow|$, which characterizes a complete decoherence and shows no difference from the Born-Markovian approximate result. In the presence of the FBS, the substitution of Eq. \eqref{longtime2} into Eq. \eqref{concurrence} results in that the state at the stroboscopic times approaches
\begin{eqnarray}
    \rho(\infty)&=&\{[\mathcal{Z}^2|\uparrow\rangle\langle \uparrow|+(1-\mathcal{Z}^2)|\downarrow\rangle\langle \downarrow|]^{\otimes 2}+|\downarrow\rangle\langle \downarrow|^{\otimes 2}\nonumber\\
    &&+[\mathcal{Z}^2\mathrm{e}^{-2\mathrm{i}\epsilon^bt/\hbar}|\uparrow\rangle \langle \downarrow|^{\otimes 2}+\text{H.c.}]\}/2.\label{ltm}
\end{eqnarray}
Choosing the concurrence $\mathcal{C}(t)=\text{max}(0,\sqrt{\lambda_1}-\sqrt{\lambda_2}-\sqrt{\lambda_3}-\sqrt{\lambda_4})$, where $\lambda_i$ is the eigenvalues of $\rho(t)\tilde{\rho}(t)$ in decreasing order and $\tilde{\rho}(t)=(\hat{\sigma}_y\otimes\hat{\sigma}_y)\rho(t)^{\ast}(\hat{\sigma}_y\otimes\hat{\sigma}_y)$, with $\hat{\sigma}_y$ being the Pauli matrix, as a measure of quantifying the entanglement \cite{PhysRevLett.80.2245}, we can calculate
\begin{equation}
    \mathcal{C}(\infty)=\mathcal{Z}^4.\label{sdtcon}
\end{equation}
Therefore, the formation of the FBS makes the entanglement partially preserved in the steady state.

%\subsection{Continuous variable system}
Second, we consider that the MQEs are formed by two magnons. As a promising solid-state platform in quantum technology \cite{10.1063/5.0020277,10.1088/1361-648X/ad6828}, the rapidly developing quantum magnonics focuses on the quantum states of magnons and their hybridization with other quantum platforms \cite{10.1016/j.physrep.2022.03.002,doi.org/10.1063/5.0157520}. However, the interaction of the magnon with the HML leads to decoherence and seriously hinders its practical application. The preservation of the magnon-magnon entanglement is particularly important for its further applications \cite{PhysRevResearch.1.023021,PhysRevB.104.224302}. We show the effect of Floquet engineering in maintaining quantum entanglement of magnons. In the framework of Feynman-Vernon's influence-functional theory in the coherent-state representation, the solution of the master equation \eqref{masterequation1} with $\hat{O}=\hat{a}$ can be written as (see Supplementary Note 1)
\begin{eqnarray}
\rho(\bar{\pmb \alpha}_{f},{\pmb \alpha}_{f}^{\prime};t)&=&\int d\mu({\pmb \alpha}_{i})d\mu({\pmb\alpha}_{i}^{\prime})\mathcal{K} (\bar{\pmb\alpha}_{f},{\pmb \alpha}_{f}^{\prime};t|\bar{\pmb \alpha}_{i}^{\prime}, {\pmb \alpha}_{i};0)\nonumber \\
&&\times\rho(\bar{\pmb \alpha}_{i},{\pmb \alpha}_{i}^{\prime};0),\label{masterequation3}
\end{eqnarray}
where $\rho(\bar{\pmb\alpha}_{f},{\pmb\alpha}_{f}^{\prime};t)=\langle\bar{\pmb\alpha}_f|\rho(t)|{\pmb\alpha}_{f}^{\prime}\rangle$ and $\rho(\bar{\pmb\alpha}_i,{\pmb\alpha}_i^{\prime};0)=\langle\bar{\pmb\alpha}_i|\rho(0)|{\pmb\alpha}_i^{\prime}\rangle$. The coherent state is defined as $|{\pmb \alpha}\rangle=\mathrm{e}^{\alpha_1\hat{a}_1^\dag+\alpha_2\hat{a}_2^\dag}|0_1,0_2\rangle$ and $d\mu({\pmb\alpha})=\prod_l \mathrm{e}^{-\bar{\alpha}_l\alpha_l}d^2\alpha_l/2\uppi$. The propagation function $\mathcal{K}(\bar{\pmb\alpha}_{f},{\pmb\alpha}_{f}^{\prime};t|\bar{\pmb\alpha}_{i},{\pmb\alpha}_{i}^{\prime};0)$ reads
\begin{eqnarray}\label{propa}
&&\mathcal{K}(\bar{\pmb\alpha}_{f},{\pmb\alpha}_{f}^{\prime};t|\bar{\pmb\alpha}_{i},{\pmb\alpha}_{i}^{\prime};0)
=\exp\{\sum_{l=1}^{2}[c(t)\bar{\alpha}_{lf}\alpha_{li} \notag \\
&&~~~~~~~~~~~~~~~+\bar{c}(t)\bar{\alpha}_{li}'\alpha_{lf}'+[1-|c(t)|^{2}]\bar{\alpha}_{li}^{\prime }\alpha _{li}\}.
\end{eqnarray}
We consider that the initial state of the magnons is a two-mode squeezed vacuum state $|\Psi(0)\rangle=\text{exp}[r(\hat{a}_{1}\hat{a}_{2}-\hat{a}^{\dagger}_{1}\hat{a}^{\dagger}_{2})]|00\rangle$. Substituting its form in the coherent-state representation $\rho(\bar{\pmb\alpha}_{i},{\pmb \alpha}_{i}';0)=\cosh^{-2} r\text{exp}[-\tanh r(\bar{\alpha}_{1i}\bar{\alpha}_{2i}+\alpha_{1i}^{\prime}\alpha_{2i}^{\prime})]$ into Eq. \eqref{masterequation3} and performing the Gaussian integration, we obtain
\begin{eqnarray}
\rho (\bar{\pmb\alpha}_{f},{\pmb\alpha}_{f}^{\prime };t) &=&o\prod_{l\neq l'}\exp (\frac{p\bar{\alpha}_{lf}\bar{\alpha}_{l'f}+\bar{p}\alpha _{lf}'\alpha _{l'f}'}{2}\nonumber\\
&&+q\bar{\alpha}_{lf}\alpha _{lf}'),\label{tdcstk}
\end{eqnarray}
where $o=\xi\cosh^{-2}r$, $p=-\xi c(t)^2\tanh r$, $q=\xi\tanh^2 r[1-|c(t)|^2]|c(t)|^2$, and $\xi=[1-(1-|c(t)|^2)^2\tanh^2 r]^{-1}$. The continuous-variable entanglement can be measured by the logarithmic negativity defined as $E_N=\max\{0,-\log_2(2\tilde{\nu}_{\text{min}})\}$ \cite{PhysRevLett.84.2726}, where $\tilde{\nu}_{\text{min}}$ being the minimal eigenvalue of the partially transposed covariance matrix $\tilde{\mathbf{V}}=\mathbf{\Lambda}\mathbf{V}\mathbf{\Lambda}$ and $\mathbf{\Lambda}=\text{diag}(1,1,1,-1)$. The element of the covariance matrix $\bf{V}$ reads $V_{mn}=\text{Tr}[(\Delta \hat{X}_{m}\Delta \hat{X}_{n}+\Delta \hat{X}_{n}\Delta \hat{X}_{m})\rho]/2$ depicted by the quadrature-operator vector $\hat{X}=(\hat{x}_1,\hat{p}_1,\hat{x}_2,\hat{p}_2)$, where $\hat{x}_l=(\hat{a}_l+\hat{a}_l^{\dagger})/\sqrt{2}$, $\hat{p}_l=(\hat{a}_l-\hat{a}_l^{\dagger})/i\sqrt{2}$, and $\Delta \hat{X}_l=\hat{X}_l-\text{Tr}(\hat{X}_{l}\rho)$. From the time-dependent state \eqref{tdcstk}, the covariance matrix can be calculated straightforwardly,
\begin{equation}
V=\left(
\begin{array}{cccc}
\frac{2n(t)+1}{2}\textbf{1}_2 & \frac{\sinh(2r)}{2}\textbf{b} \\
\frac{\sinh(2r)}{2}\textbf{b} & \frac{2n(t)+1}{2}\textbf{1}_2 \end{array}\right),
\end{equation}
where $\textbf{b}=\left(
\begin{array}{cccc}
-\text{Re}c(t)^2 & -\text{Im}c(t)^2 \\
-\text{Im}c(t)^2 & \text{Re}c(t)^2 \end{array}\right)$,
and $n(t)=|c(t)|^2\sinh^2{r}$. The logarithmic negativity $E_N(t)$ can then be calculated. It is easy to verify that the initial entanglement is $E_N(0) = \frac{2r}{\ln 2}$. In the absence of the FBS, $c(\infty)=0$ readily leads to $\rho (\bar{\pmb\alpha}_{f},{\pmb\alpha}_{f}^{\prime };\infty)=1$, which corresponds to $\rho(\infty)=|00\rangle\langle00|$ and denotes a complete decoherence. In the presence of the FBS, we can use Eq. \eqref{longtime2} to evaluate the steady-state entanglement at the stroboscopic times as
\begin{equation}
    E_N(\infty)=-\log_2 |1-\mathcal{Z}^2(1-\mathrm{e}^{-2r})|,\label{lgngt}
\end{equation}
which indicates that the formation of the FBS overcomes the degradation of the entanglement of the continuous-variable system under local decoherence. We find that the dominated role played by the single-excitation quasi-energy spectrum of the total system in the reduced dynamics of the periodically driven system is still valid for the magnon system (see Supplementary Note 1).

\subsection*{Numerical results}\label{entanglement preservation}

\begin{figure}[tbp]
\centering
\includegraphics[width=\columnwidth]{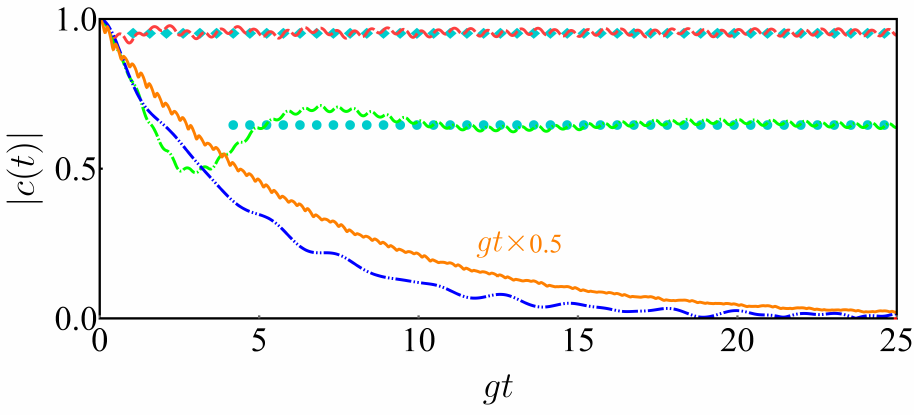}
\caption{\textbf{Fidelity evolution in the cases with and without periodic driving.} Evolution of fidelity $|c(t)|$ within the driving amplitudes $\mathcal{F}=0$ (blue dot-dot-dashed line), $5g$ (green dot-dashed line), $10g$ (red dashed line), and $22g$ (orange solid line). The actual evolution time of the orange line is twice that of other cases. The cyan dots and squares are their corresponding stroboscopic dynamics \eqref{longtime2} at integer multiples of period $T$ evaluated by the Floquet approach. We use the driving frequency $\Omega=12g$, the driving duration $t^{\prime}=T/2$, the eigen-frequency of magnetic quantum entity (MQE) $\omega_0=g$,  the center frequency of hybrid magnetic lattice (HML) $\omega_c=0.5g$, the nearest-neighbor coupling strength between magnetic particles in HML $h=1.5g$, and the size of HML $N=800$. We use the coupling strength $g$ between HML and MQE as the frequency scale.}\label{dfdfd}
\end{figure}

\begin{figure}
  \centering
  \includegraphics[width=\columnwidth]{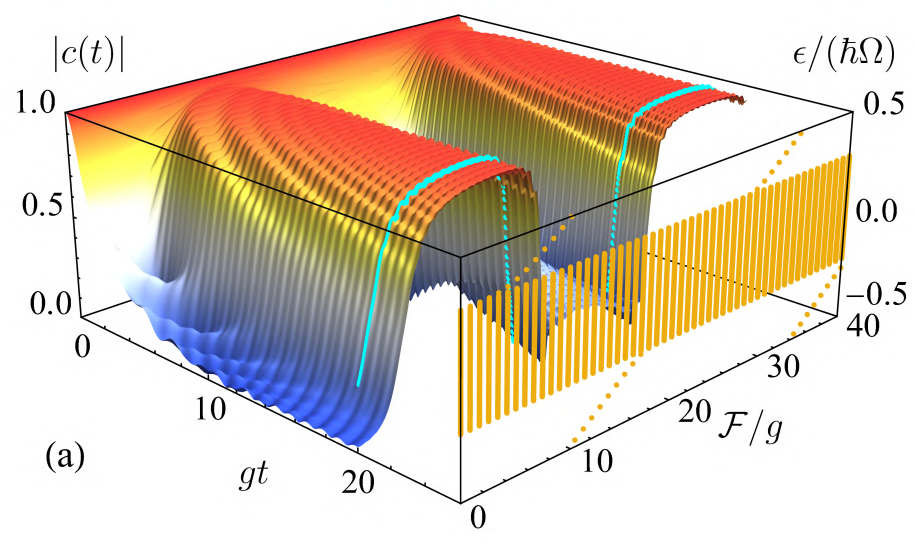}\\
  \includegraphics[width=\columnwidth]{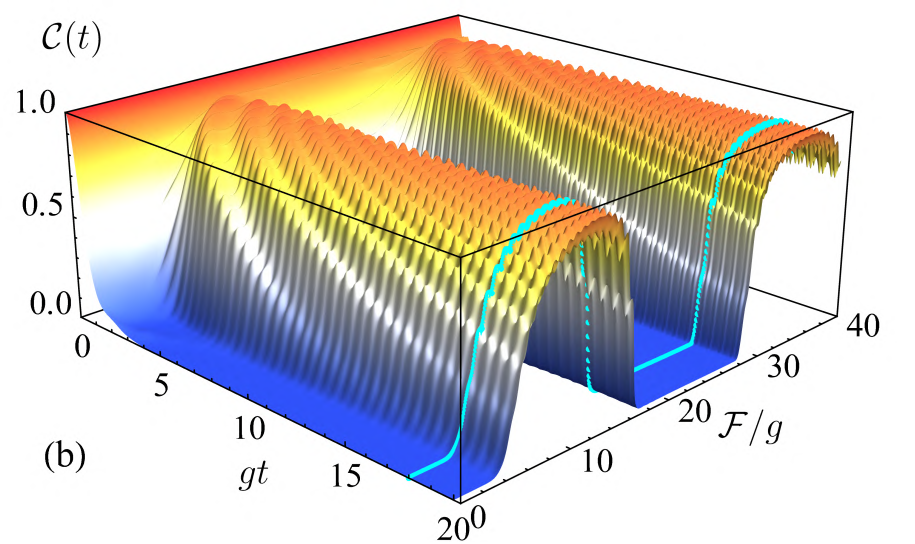}
  \caption{\textbf{The quasi-energy spectrum and dynamic evolution of fidelity and discrete-variable concurrence.} (\textbf{a}) The quasi-energy spectrum and the evolution of fidelity $|c(t)|$ as a function of driving amplitude $\mathcal{F}$. (\textbf{b}) The evolution of concurrence $\mathcal{C}(t)$ between two nitrogen-vacancy centers. The long-time values of $|c(t)|$ evaluated from Eq. \eqref{longtime2} and $\mathcal{C}(t)$  evaluated from Eq. \eqref{sdtcon} at $t=35T$ are represented by the cyan dotted lines, respectively. The yellow dots in (\textbf{a}) show the quasienergy spectrum, which consists of a continuous energy band and one possibly formed FBS in the gap. Other parameters are the same as Fig. \ref{dfdfd}.}\label{Fig2}
\end{figure}
\begin{figure}
  \includegraphics[width=\columnwidth]{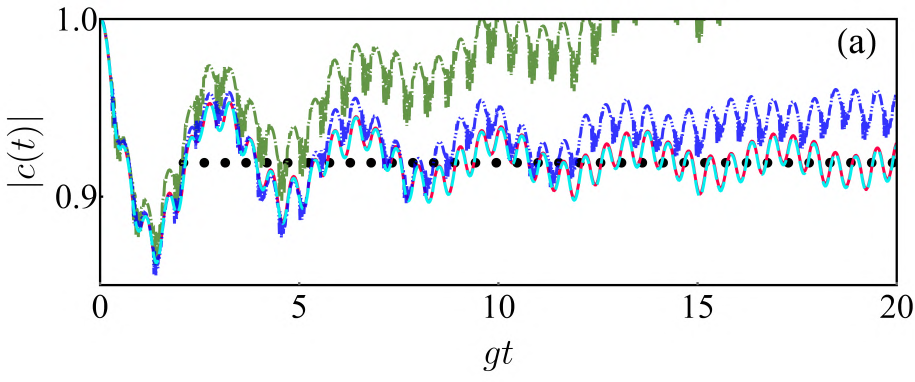}\\
  \includegraphics[width=\columnwidth]{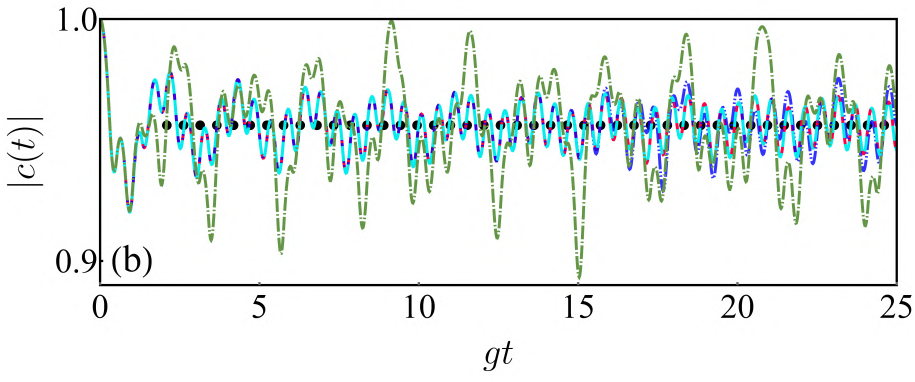}
  \caption{\textbf{Influences of the time step and the finite size effect on the dynamic evolution of fidelity.} (\textbf{a}) Evolution of fidelity $|c(t)|$ with the driving amplitude $\mathcal{F}=8g$ and the time step to evaluate Eq. \eqref{dynamics1} at $\Delta t=0.02g^{-1}$ (green dot-dashed line), 0.017$g^{-1}$ (blue dot-dot-dashed line), 0.002$g^{-1}$ (red dashed line), 0.0004$g^{-1}$ (cyan solid line). (\textbf{b}) Evolution of fidelity $|c(t)|$ with the driving amplitude $\mathcal{F}=10g$ and the size of HML is chosen as $N=5$ (green dot-dashed line), 50 (blue dot-dot-dashed line), 200 (red dashed line), 500 (cyan solid line). The black dots represent their corresponding stroboscopic dynamics from Eq. \eqref{longtime2} at integer multiples of period $T$ evaluated by the Floquet approach. Other parameters are the same as Fig. \ref{dfdfd}.}\label{convg}
\end{figure}

\begin{figure}
\centering
\includegraphics[width=\columnwidth]{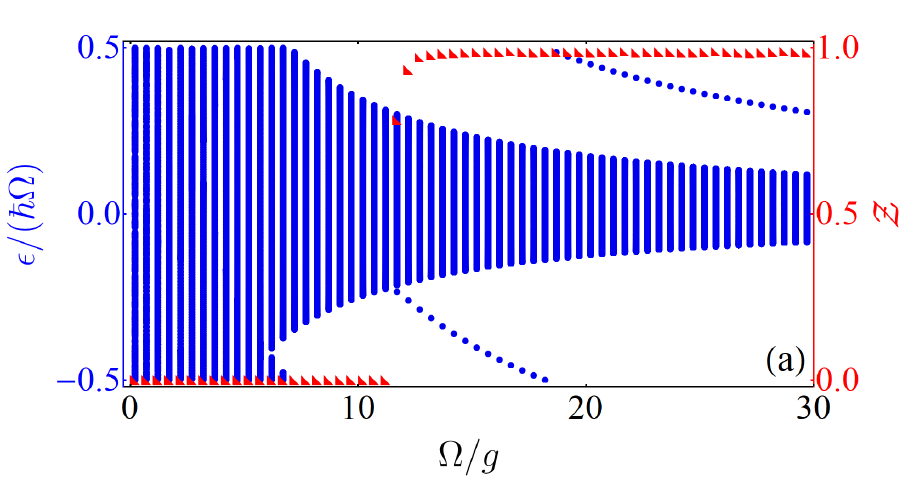}\\
\includegraphics[width=\columnwidth]{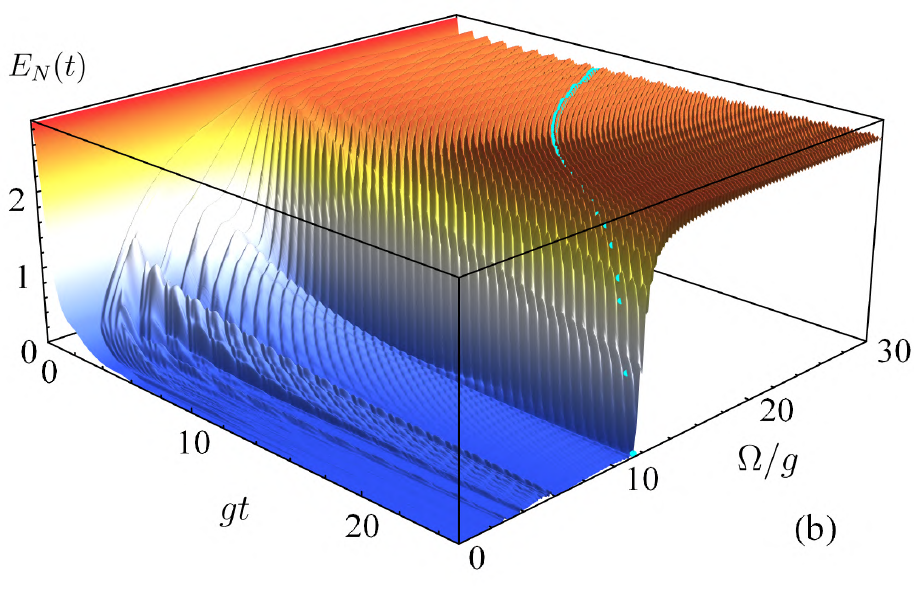}\\
\caption{\textbf{The quasi-energy spectrum and the dynamic evolution of continuous-variable logarithmic negativity.} (\textbf{a}) The quasienergy spectrum $\epsilon$ (blue dots) and the occupation probability  $\mathcal{Z}$  (red triangle) and (\textbf{b}) the evolution of logarithmic negativity $E_N(t)$ between two magnons as a function of driving frequency $\Omega$. The long-time behavior of $E_N(t)$ at $t=45T$ from Eq. \eqref{lgngt} is shown by the cyan dotted line. Other parameters are the same as Fig. \ref{Fig2}, except for driving amplitude $\mathcal{F}=16g$ and squeezing parameter $r=1$.}\label{Fig3}
\end{figure}

We perform numerical calculations to verify our analytical result. Figure \ref{dfdfd} shows the evolution of $|c(t)|$ obtained by numerically solving Eq. \eqref{dynamics1}. In the static limit of $\mathcal{F}=0$, $|c(t)|$ completely decays to zero and signifies a complete decoherence. The vanishing fate of $|c(t)|$ can be changed when the periodic driving is turned on. Manifesting a clear decoherence suppression, such a behavior is absent in the Markovian approximate case. We find that the stabilized values of $|c(t)|$ with tiny-amplitude oscillation matches well the values evaluated from the analytical equation \eqref{longtime2}. To gain a global picture of the parameter regimes of the periodic driving to suppress the decoherence, we plot in Fig. \ref{Fig2}(a) the evolution of $|c(t)|$ in different values of the driving amplitude $\mathcal{F}$. In the regimes $\mathcal{F}\in (3g,18g)$ and $(28g,40g)$, the behavior of damping to zero of $|c(t)|$ is completely overcome and $|c(t)|$ approaches a lossless oscillation. We find that the regions where the decoherence is suppressed match exactly with the regions where the FBS is formed in the quasienergy spectrum of $\hat{H}_{l}(t)$. The result proves the dominated role of the FBS in the steady state of the periodically driven NV centers. It is further confirmed by the coincidence of the long-time values of $|c(t)|$ at the stroboscopic times obtained from the analytical result in Eq. \eqref{longtime2} with the long-time values of $|c(t)|$ obtained by numerically solving Eq. \eqref{dynamics1}, see the cyan dotted line in Fig. \ref{Fig2}(a). Choosing the MQEs as NV centers, we plot in Fig. \ref{Fig2}(b) the evolution of the concurrence for the initial maximally entangled state of the two NV centers. We find that as long as the damping to zero of $|c(t)|$ is suppressed due to the formation of the FBS, a considerable amount of entanglement between the NV centers is persistently preserved in the steady state. The values of the preserved entanglement at the stroboscopic times agree well with our analytical result in Eq. \eqref{sdtcon}, see the cyan dotted line in Fig. \ref{Fig2}(b). All the results demonstrate the distinguished role of the FBS in suppressing the decoherence of the NV centers.

It should be noted that our numerical results are convergent. Their correction is ensured by the fact that all the numerical results in the long-time limit converge to the analytical results evaluated by the Floquet approach. To demonstrate this point, we plot in Fig. \ref{convg}(a) the evolution of $|c(t)|$ in different values of the time step $\Delta t$ to numerically calculating Eq. \eqref{dynamics1}. We indeed see that when $\Delta t=0.02g^{-1}$ and $0.017g^{-1}$, the results have a large deviation from the analytical results at stroboscopic time $t=mT$ evaluated from the Floquet approach. When $\Delta t=0.002g^{-1}$, the numerical results begin to match the analytical results. With further decreasing $\Delta t$, the numerical results do not change anymore. It means that our numerical results when $\Delta t=0.002g^{-1}$ have reached convergence. We also plot in Fig. \ref{convg}(b) the evolution of $|c(t)|$ in different size $N$ of HML. We find that when $N=5$, the numerical result cannot match the analytical result at stroboscopic time. When $N=200$, the numerical result matches the analytical result only when $ t<15g^{-1}$ [see the red line in Fig. \ref{convg}(b)]. When $N=500$, the numerical and analytical results match well within our interested time interval. Therefore, as long as the convergence is not achieved, the numerical result would deviate from the analytical one.

\begin{figure}[tbp]
\centering
\includegraphics[width=\columnwidth]{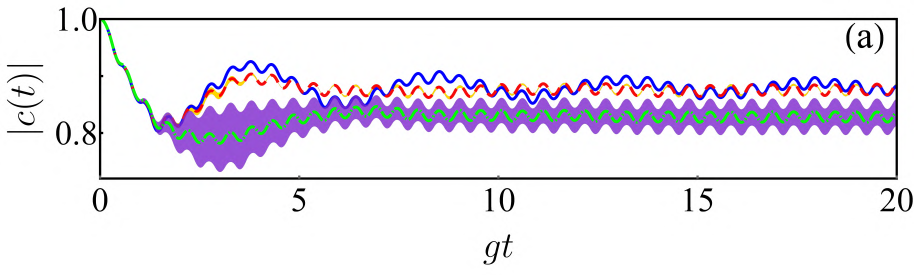}\\
\includegraphics[width=\columnwidth]{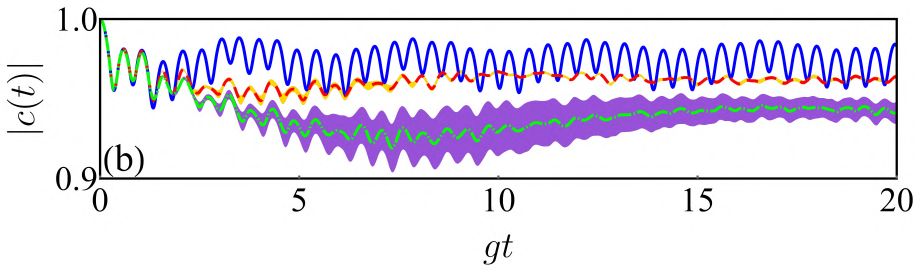}\\
\includegraphics[width=\columnwidth]{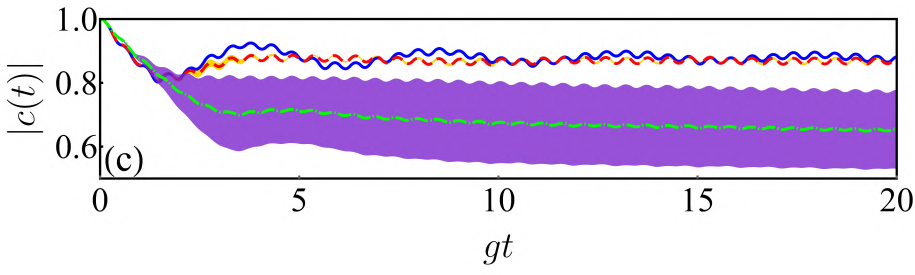}
\caption{\textbf{Influences of the fluctuations on the dynamic evolution of fidelity.} Evolution of fidelity $|c(t)|$ when a fluctuation $w\chi$ is added on the driving amplitude $\mathcal{F}$ in (\textbf{a}), on the driving frequency $\Omega$ in (\textbf{b}), and on the eigen-frequency $\omega_0$ in (\textbf{c}), respectively. The red dashed and green dot-dashed lines are the average values obtained by averaging over 100 random configurations with the fluctuation amplitude $w=1.2g$ and $3g$ in (\textbf{a}), $w=1.2g$ and $1.5g$ in (\textbf{b}), and $w=0.8g$ and $8g$ in (\textbf{c}), respectively. The orange and purple regions are their respective variations. The blue solid lines are the fluctuation-free results. We chose the driving amplitude and frequency as $(\mathcal{F},\Omega)=(7g,12g)$ in (\textbf{a}) and (\textbf{c}) and $(16g,13g)$ in (\textbf{b}). Other parameters are the same as Fig. \ref{Fig2}. }\label{FigR4}
\end{figure}

The quasienergy spectrum in different values of the driving frequency $\Omega$ is given in Fig. \ref{Fig3}(a). The result reveals that when $\Omega<4h$, the quasienergy spectrum is filled by the continuous band and there is no room for forming the FBS. This can be understood as follows. Being periodic with a period $\hbar\Omega$, the quasienergy has a full width $\hbar\Omega$. The bandwidth of the HML is $4\hbar h$. Therefore, only when $\Omega>4h$, a band gap with a width $\hbar(\Omega-4h)$ can be present in the quasienergy spectrum. Therefore, $\Omega>4h$ is a necessary condition for the driving field to suppress the decoherence. Figure \ref{Fig3}(a) also indicates that a FBS is formed when $\Omega$ further increases to $12g$. Whenever the FBS is formed, $|c(t)|$ tends to a nonzero value $\mathcal{Z}$ given by Fig. \ref{Fig3}(a). When the MQEs consist of two magnon modes, the evolution of the entanglement between two magnons in Fig. \ref{Fig3}(b) confirms again that, accompanying with the formation of the FBS, the damping of the entanglement is completely avoided and the entanglement tends to a lossless oscillation in a tiny amplitude. The steady-state value of the entanglement at the stroboscopic times matches well with our analytical result in Eq. \eqref{lgngt}. It verifies again that the entanglement preservation in our exact non-Markovian dynamics is caused by the formation of the FBS.

To reveal the impact of the fluctuation of driving parameters on decoherence suppression, we add a random fluctuation $w\chi$, with $\chi$ being a random number uniformly distributed in $[-1,1]$ and $w$ being the fluctuation strength, on the driving amplitude $\mathcal{F}$ and frequency $\Omega$. Figure \ref{FigR4} shows the evolution of the mean values of $|c(t)|$ obtained by averaging 100 random fluctuations and their variances. We can see from Fig. \ref{FigR4}(a) that when the fluctuation of the driving amplitude has a strength $w$ as high as $3g$, the mean values of $|c(t)|$ still show little difference from the fluctuation-free results. Even the driving frequency $\Omega$ experiences a fluctuation with $w=1.5g$ in Fig. \ref{FigR4}(b), $|c(t)|$ can still be stabilized. It clearly demonstrates the robustness of our Floquet-engineering scheme to parameter fluctuation in suppressing the decoherence.

\begin{figure}[tbp]
\centering
\includegraphics[width=\columnwidth]{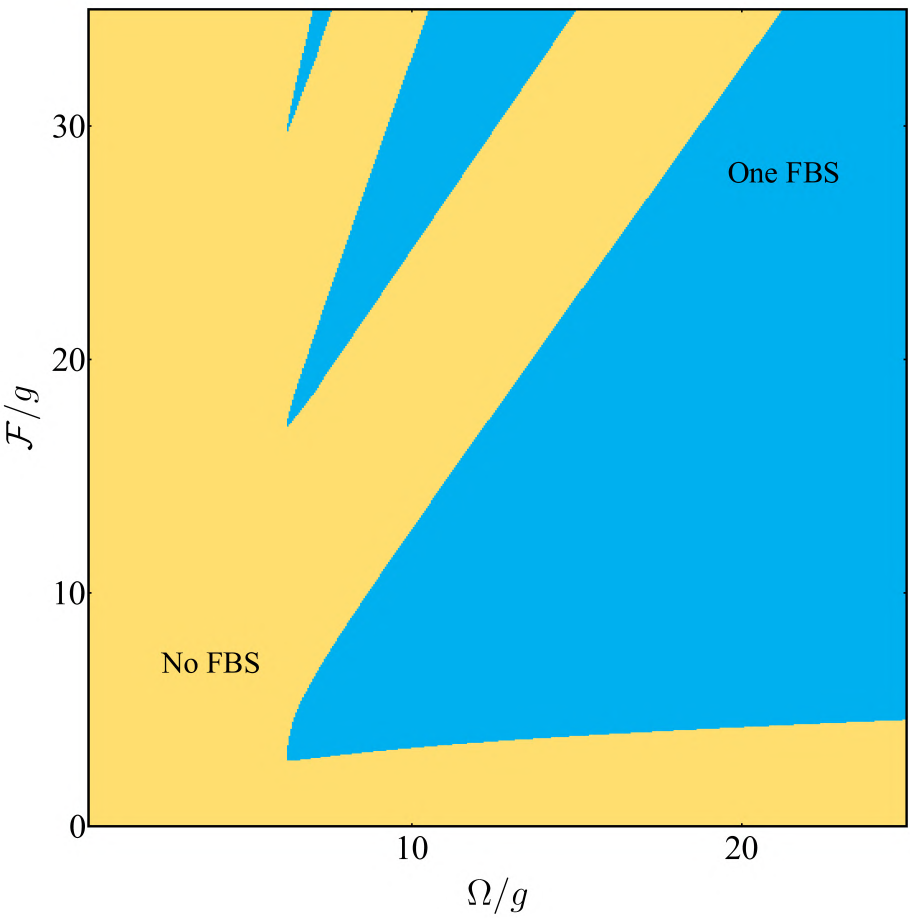}%\columnwidth
\caption{\textbf{Phase diagram characterized by whether the Floquet bound state is present or not in the quasi-energy spectrum.} One Floquet bound state (FBS) exists in the blue parameter region, while no FBS exists in the yellow region. Other parameters are the same as Fig. \ref{Fig2}. }\label{pdfrb}
\end{figure}

In practice, the MQEs generally feel the dephasing noises, which result in the expansion of their frequency $\omega_0$. In order to explore the effect of dephasing noise on our scheme, we add a random static energy shift $w\chi$ on $\hbar\omega_0$. Figure \ref{FigR4}(c) shows the obtained results. This indicates that our scheme still exhibits a good decoherence-suppression performance even when the MQEs experience a random energy shift with $w=8g$. Such a robustness of our scheme can be understood as follows. Our Floquet-engineering scheme for suppressing the decoherence is based on the formation of a FBS. We can find that the quasienergy spectrum in the presence of the FBS contains several wide regions instead of certain single values of the driving parameters. This makes our scheme robust to the imperfect fluctuation of the driving parameters. Therefore, the efficiency of the scheme in suppressing the decoherence of MQEs caused by the HMLs is protected by the feature of the quasienergy spectrum of the total MQE-HML system. To gain a global picture, we plot in Fig. \ref{pdfrb} the phase diagram characterized by whether the FBS is present or not in the parameter space of the driving amplitude and frequency. This provides a useful guide for choosing the driving parameters.

The above analysis reveals an insightful mechanism to control quantum states in hybrid magnetic systems for both the discrete- and continuous-variable cases. It indicates that one can suppress the decoherence of the MQEs induced by the HMLs by actively applying a periodic driving on the system such that a FBS is formed in the quasienergy spectrum of the total MQE-HML system. This gives a useful instruction for designing magnetic quantum networks \cite{10.1126/science.abg1919,PhysRevLett.127.183202,PRXQuantum.2.040344}.

\section*{Discussion}\label{Discussion}
Compared to the dynamical decoupling method \cite{PhysRevLett.82.2417, PhysRevA.58.2733}, our Floquet engineering approach does not require high-frequency pulses or precise timing sequences. It makes our scheme more experimentally accessible. Moreover, the dynamical decoupling method is known to be equivalent to the spectral filtering method \cite{PhysRevLett.82.2417,PhysRevA.58.2733,PhysRevLett.87.270405}. It has been shown that the spectral filtering method breaks down in the strong-coupling regime, while the Floquet engineering method can still accurately describe the long-time dynamics of periodically driven open systems \cite{PhysRevA.91.052122}. This further highlights the advantage of our Floquet engineering approach.
Our scheme is realizable in state-of-the-art experiments. On the one hand, the fabrication of YIG in nanoscale and the observation of strong coupling between magnon modes in a YIG sphere with magnetic qubits and superconducting coplanar microwave resonators have been experimentally realized \cite{PhysRevLett.123.107702,doi.org/10.1007/s12274-020-3251-5,PhysRevLett.113.083603,PhysRevLett.111.127003}, which is helpful for the realization of our hybrid HML system. On the other hand, the eigen-frequency of the MQE composed of either NV centers or magnon modes is determined by the magnetic field. It is easy to control by introducing auxiliary periodic drivings \cite{PhysRevA.103.043706,PhysRevResearch.4.L012025}. In our system, with structure parameters $(R, l_0, d)\sim$ nm, $\omega_{\pm} \sim 10^{9}$ Hz, and $(B_0, B_z,B_s)\sim$ T, the eigenfrequency of MQE $\omega_0$, the center frequency $\omega_c$, the coupling strength $g$ and the loop-mediated interaction $h$ in our calculations are in the order of $10^9$ Hz \cite{PhysRevA.100.022343,PhysRevLett.125.153602,PhysRevA.103.043706,PhysRevB.94.224410,PhysRevB.108.224416,PhysRevLett.111.127003}. Our numerical simulations use $g$ as the frequency scale. Therefore, it is only the relative values of other parameters to $g$ that play the dominated role in our Floquet-engineering scheme. In different MQE frequency regimes, the quantitative condition for forming the FBS might explicitly change, but the hidden mechanism that the formation of the FBS suppresses the dissipation of the MQEs caused by the HMLs does not change. Note that our scheme to suppress the decoherence of MQEs exerted by the HMLs in the discrete-variable case is applicable in both a magnetic dipolar emitter \cite{PhysRevLett.125.247702,PhysRevB.108.L180409} and a spin-1/2 defect \cite{Yang_2019,PhysRevResearch.4.043180} as the two-level systems. Floquet engineering has been effectively applied in hybrid magnetic systems, supporting the feasibility of our Floquet scheme \cite{PhysRevX.12.021061,PhysRevLett.125.237201,PhysRevApplied.23.024053,wc1j-mq69,PhysRevApplied.22.054048,PhysRevApplied.14.054033,PhysRevA.110.013711}.

\section*{Conclusions}\label{conclusion}
In conclusion, we have studied the non-Markovian dynamics of two MQEs, which are formed by either NV centers or magnon modes, coupled to two independent HMLs. We have proposed a Floquet engineering scheme to suppress the decoherence of the MQEs caused by the HMLs. It is found that, as long as a FBS is present in the quasienergy spectrum of the total system of each periodically driven MQE and its HML, the decoherence is overcome and a persistent entanglement between the MQEs is kept in the steady state. This mechanism provides a universal approach to decoherence suppression: once the condition for forming a Floquet bound state is met, the suppression itself is robust and independent of the specific value of the magnetic coupling strength $g$. Solving the decoherence problem of MQEs in HML, our result is helpful for the application of the HML in quantum technologies.

\section*{Funding Declaration}
This work is supported by the National Natural Science Foundation of China (Grants No. 12275109, No. 92576202, No. 12074106, and No. 12247101), the Innovation Program for Quantum Science and Technology of China (Grants No. 2023ZD0300904 and No. 2023ZD0300400), the Natural Science Foundation of Henan Province (Grant No. 242300421165 and No. 252300421779), and the Hubei Province Science Fund for Distinguished Young Scholars (Grant No. 2020CFA078).

\section*{Data availability}

The numerical data for generating the figures are available from the corresponding author (Jun-Hong An) upon request.

\section*{Code availability}

The numerical codes for generating the results are also available from the corresponding author (Jun-Hong An) upon request.

\section*{Author Contributions}

Jun-Hong An proposed the original idea and developed the theoretical formulism. Feng-Zhou Ji, Si-Yuan Bai performed the analytic derivations. Feng-Zhou Ji drafted the manuscript and designed the figures. Feng-Zhou Ji, Si-Yuan Bai, Wan-Li Yang, Chun-Jie Yang, and Jun-Hong An discussed the results and revised the manuscript.

\section*{Competing interests}

The authors declare no competing interests.

\bibliography{Ref}

%\section*{References}

\end{document}

\title{Supplementary Information for ``Floquet engineering in hybrid magnetic quantum systems''}
\author{Feng-Zhou Ji\orcidlink{0000-0003-4859-1535}}
\affiliation{School of Physics, Henan Normal University, Xinxiang 453007, China}
\author{Si-Yuan Bai\orcidlink{0000-0002-4768-6260}}
\affiliation{Key Laboratory of Quantum Theory and Applications of MoE, Lanzhou Center for Theoretical Physics, Key Laboratory of Theoretical Physics of Gansu Province, Gansu Provincial Research Center for Basic Disciplines of Quantum Physics, Lanzhou University, Lanzhou 730000, China}
\author{Wan-Li Yang}
\affiliation{State Key Laboratory of Magnetic Resonance and Atomic and Molecular Physics, Innovation Academy for Precision Measurement Science and Technology, Chinese Academy of Sciences, Wuhan 430071, China}
\author{Chun-Jie Yang\orcidlink{0000-0003-2137-6958}}
\email{yangchunjie@htu.edu.cn}
\affiliation{School of Physics, Henan Normal University, Xinxiang 453007, China}
\author{Jun-Hong An\orcidlink{0000-0002-3475-0729}}
\email{anjhong@lzu.edu.cn}
\affiliation{Key Laboratory of Quantum Theory and Applications of MoE, Lanzhou Center for Theoretical Physics, Key Laboratory of Theoretical Physics of Gansu Province, Gansu Provincial Research Center for Basic Disciplines of Quantum Physics, Lanzhou University, Lanzhou 730000, China}

\maketitle

\section*{Supplementary Note 1: The derivation of non-Markovian master equation (6) and its solutions}
We give the detailed derivation of the non-Markovian master equation (6) in the main text and its solutions in Eqs. (14) and (19) for the discrete- and continuous-variable systems, respectively. Our system consists of two sets of subsystems. Each subsystem consists of a
periodically driven MQE coupled to a one-dimensional HML. The Hamiltonian is $\hat{H}%
(t)=\sum_{l=1}^{2}\hat{H}_{l}(t)$, with $\hat{H}_{l}(t)=\hat{H}_{l}^{\text{%
sys}}(t)+\hat{H}_{l}^{\text{hml}}+\hat{H}_{l}^{\text{int}}$ and
\begin{eqnarray}
\hat{H}_{l}^{\text{sys}}(t) &=&\hbar \lbrack \omega _{0}+A(t)]\hat{O}%
_{l}^{\dag }\hat{O}_{l}, \\
\hat{H}_{l}^{\text{hml}} &=&\sum_{k}\hbar \omega _{k}\hat{\tilde{f}}_{k}^{\dag l}\hat{\tilde{f}}_{k}^{l}, \\
\hat{H}_{l}^{\text{int}} &=&-\hbar \sum_{k}g_{k}(\hat{O}_{l}^{\dag }\hat{\tilde{f}}%
_{k}^{l}+\text{H.c.}).
\end{eqnarray}%
Because there is no interaction between the two subsystems, the two subsystems evolve with time independently and the two HMLs act as two environments locally influencing the two MQEs. The HMLs cause the dissipative decoherence to the two MQEs. Under the condition that the initial states of the two HMLs are vacuum state $\otimes _{l=1}^{2}|\{0_{k}\}_{l}\rangle $, we can derive the exact master equation of the MQEs for both the discrete- and continuous-variable cases from the unitary dynamics of total system via tracing out the degrees of freedom of the HMLs.

\subsection*{Discrete-variable system}\label{dvsdr}

In this case, the MQEs are formed by two NV centers described by two-level systems. Thus, we have $\hat{O}_{l}=\hat{\sigma}_{l}$, where $\hat{\sigma}_{l}=|\downarrow_{l}\rangle \langle \uparrow_{l}|$ is the transition operator of the $l$th NV center from its excited state $|\uparrow_{l}\rangle $ to its ground state $|\downarrow_{l}\rangle $. First, we consider the following two special cases of the initial state of the $l$th subsystem.

\begin{enumerate}
\item The initial state is $|\Psi _{1}(0)\rangle=|\downarrow_{l},\{0_{k}\}_{l}\rangle $. Since it is an eigenstate of $\hat{H}_{l}$ with eigenenergy zero, it does not change with time. Thus, we have $|\Psi_{1}(t)\rangle =|\Psi _{1}(0)\rangle $.

\item The initial state is $|\Psi _{2}(0)\rangle=|\uparrow_{l},\{0_{k}\}_{l}\rangle $. Because the total exitation number $\hat{\mathcal N}=\hat{\sigma}^\dag_l\hat{\sigma}_l+\sum_k\hat{\tilde f}_k^{\dag l}\hat{\tilde f}_k^l$ is conserved, the evolved state from  $|\Psi _{2}(0)\rangle$ can be expanded as
\begin{equation}
|\Psi _{2}(t)\rangle =c(t)|\uparrow_{l},\{0_{k}\}_{l}\rangle+\sum_{k}d_{k}(t)|\downarrow_{l},1_{k,l}\rangle ,\label{apptmp2}
\end{equation}%
where $c(t)$ is the probability amplitude of $l$th NV center in the excited state and $|1_{k,l}\rangle$ denotes the state of the $l$th HML with only one excitation in its $k$th mode. From the time-dependent Schr\"{o}dinger equation $i\hbar|\dot{\Psi}(t)\rangle=\hat{H}_l|\Psi(t)\rangle$, we obtain 
\begin{eqnarray}
\mathrm{i}\dot{c}(t) &=&[\omega _{0}+A(t)]c(t)-\sum_{k}g_{k}d_{k}(t),  \label{xc} \\
\mathrm{i}\dot{d}_{k}(t) &=&\omega _{k}d_{k}(t)-g_{k}c(t).  \label{dd}
\end{eqnarray}
Substituting the solution of Eq. (\ref{dd}) under the initial condition $d_k(0)=0$ as $d_{k}(t)=\mathrm{i}g_{k}\int_{0}^{t}d\tau \mathrm{e}^{-\mathrm{i}\omega _{k}(t-\tau )}c(\tau )$ into Eq. (\ref{xc}), we have
\begin{equation}
\dot{c}(t)+\mathrm{i}[\omega _{0}+A(t)]c(t)+\int_{0}^{t}d\tau F(t-\tau )c(\tau )=0,\label{cctd}
\end{equation}
under the initial condition $c(0)=1$, where $F(t-\tau )=\int d\omega J(\omega )\mathrm{e}^{-\mathrm{i}\omega (t-\tau )}$ and $J(\omega )=\sum_{k}g_{k}^{2}\delta (\omega -\omega _{k})$ are the correlation function and the spectral density of the HML, respectively. 
\end{enumerate}

With these two special cases at hand, the evolution of an arbitrary initial state of the $l$th subsystem $\rho _{\text{tot}}(0)=\rho _{l}(0)\otimes|\{0_{k}\}_{l}\rangle \langle \{0_{k}\}_{l}|$, where $\rho _{l}(0)=\rho_{ee}|\uparrow_{l}\rangle \langle \uparrow_{l}|+\rho _{gg}|\downarrow_{l}\rangle \langle \downarrow_{l}|+(\rho _{eg}|\uparrow_{l}\rangle \langle \downarrow_{l}|+\text{h.c.})$, is readily obtained as
\begin{eqnarray}
\rho _{\text{tot}}(t) &=&\rho _{ee}|\Psi _{2}(t)\rangle \langle \Psi_{2}(t)|+\rho _{gg}|\downarrow_{l},\{0_{k}\}_{l}\rangle \langle \downarrow_{l},\{0_{k}\}_{l}| \nonumber\\
&&+[\rho _{eg}|\Psi _{2}(t)\rangle \langle \downarrow_{l},\{0_{k}\}_{l}|+\text{h.c.}].
\end{eqnarray}
After tracing over the degrees of freedom of the HML, we have
\begin{eqnarray}
\rho _{l}(t) &=&\text{Tr}_{\text{hml}}[\rho _{\text{tot}}(t)] =\rho _{ee}|c(t)|^{2}|\uparrow_{l}\rangle \langle \uparrow_{l}|+(1-\rho_{ee}|c(t)|^{2}) \nonumber\\
&&  \times|\downarrow_{l}\rangle \langle \downarrow_{l}|+[\rho _{eg}c(t)|\uparrow_{l}\rangle \langle \downarrow_{l}|+\text{h.c.}],  \label{smrdcd}
\end{eqnarray}%
where $\rho _{ee}+\rho _{gg}=1$ has been used. In the basis formed by $%
|\uparrow_{l}\rangle $ and $|\downarrow_{l}\rangle $, its time derivative is 
\begin{eqnarray}
\dot{\rho}_{l}(t) &=&\Gamma (t)\left( 
\begin{array}{cc}
-2\rho _{ee}|c(t)|^{2} & -\rho _{eg}c(t) \\ 
-\rho _{ge}c^{\ast }(t) & 2\rho _{ee}|c(t)|^{2}%
\end{array}%
\right)   \nonumber\\
&& -\mathrm{i}\Omega (t)\left( 
\begin{array}{cc}
0 & \rho _{eg}c(t) \\ 
-\rho _{ge}c^{\ast }(t) & 0%
\end{array}%
\right) ,  \label{densitymatrix2}
\end{eqnarray}%
where $\Gamma (t)=-\text{Re}[\frac{\dot{c}(t)}{c(t)}]$ and $\Omega (t)=-\text{Im}[\frac{\dot{c}(t)}{c(t)}]$ denote the decay rate and renormalized frequency of the quantum system. Using Eq. \eqref{smrdcd} to rewrite $\left( 
\begin{array}{cc}
0 & \rho _{eg}c(t) \\ 
-\rho _{ge}c^{\ast }(t) & 0%
\end{array}%
\right) =[\hat{\sigma}_{l}^{\dagger }\hat{\sigma}_{l},\rho_l (t)]$ and $\left( 
\begin{array}{cc}
-2\rho _{ee}|c(t)|^{2} & -\rho _{eg}c(t) \\ 
-\rho _{ge}c^{\ast }(t) & 2\rho _{ee}|c(t)|^{2}%
\end{array}%
\right) =2\hat{\sigma}_{l}\rho_l (t)\hat{\sigma}_{l}^{\dagger }-\hat{\sigma}%
_{l}^{\dagger }\hat{\sigma}_{l}\rho_l (t)-\rho_l (t)\hat{\sigma}_{l}^{\dagger }%
\hat{\sigma}_{l}\equiv \check{\mathcal{L}}_{l}\rho_l(t)$, we obtain the exact master equation of the $l$th NV center
as 
\begin{equation}
\dot{\rho}_{l}(t)=-\mathrm{i}\Omega (t)[\hat{\sigma}_{l}^{\dag }\hat{%
\sigma}_{l},\rho _{l}(t)]+\Gamma (t)\check{\mathcal{L}}_{l}\rho (t). \label{smexactME}
\end{equation}
Based on the fact the two subsystems evolve with time independently, we thus have the master equation of two NV centers as
\begin{equation}
\dot{\rho}(t)=\sum_{l=1}^{2}\{-\mathrm{i}\Omega (t)[\hat{\sigma}_{l}^{\dag }\hat{%
\sigma}_{l},\rho(t)]+\Gamma (t)\check{\mathcal{L}}_{l}\rho (t)\}.\label{sddaa}
\end{equation}
Therefore, the master equation \eqref{sddaa} is exact under the condition that the two HMLs are in the vacuum states initially. The initial state of the two NV centers is $\rho(0)={1\over 2}\{(|\uparrow\rangle\langle \uparrow|)^{\otimes 2}+(|\downarrow\rangle\langle \downarrow|)^{\otimes 2}+[(|\uparrow\rangle\langle \downarrow|)^{\otimes 2}+\text{H.c.}]\}$. Solving Eq. \eqref{sddaa} or equivalently using Eq. \eqref{smrdcd} under this initial condition, we obtain
\begin{eqnarray}
\rho (t) &=&\{[P_{t}|\uparrow\rangle \langle \uparrow|+(1-P_{t})|\downarrow\rangle \langle \downarrow|]^{\otimes 2}+|\downarrow\rangle \langle \downarrow|^{\otimes 2}  \nonumber\\
&&+[c^{2}(t)|\uparrow\rangle \langle \downarrow|^{\otimes 2}+\text{H.c.}]\}/2,
\end{eqnarray}
where $P_{t}=|c(t)|^{2}$. 

\subsection*{Continuous-variable system} \label{cvsdrv}
In this case, the MQEs are formed by two magnons described by harmonic oscillators. Thus, we have $\hat{O}_l=\hat{a}_l$, where $\hat{a}_l$ is the annihilation operator of the $l$th magnon. The non-Markovian master equation in this case under the initial condition $\rho_\text{tot}(0)=\rho(0)\otimes|\{0_k\}_1,\{0_k\}_2\rangle\langle\{0_k\}_1,\{0_k\}_2|$ can be derived by the Feynman-Vernon's influence functional method in the coherent-state representation \cite{PhysRevA.76.042127,AN20091737}.
\subsubsection*{The path-integral influence functional method}
The evolution of the state of the total system formed by the $l$th magnon and its HML is
\begin{equation}
    \rho_{\text{tot}}^{(l)}(t)=\hat{U}_l(t)\rho_{\text{tot}}^{(l)}(0)\hat{U}^\dag_l(t),
\end{equation}
where $\hat{U}_l(t)=\mathcal{T}\exp[-{\mathrm{i}\over\hbar}\int_0^t\hat{H}_l(\tau)d\tau]$. Introducing the representation of the (unnormalized) coherent state of the magnon $|\alpha\rangle=\mathrm{e}^{\alpha\hat{a}_l^\dag}|0\rangle$ and the HML $|{\bf z}\rangle=\prod_k\mathrm{e}^{z\hat{\tilde f}^{\dag l}_k}|\{0_k\}_l\rangle$, which satisfy $\int \mathrm{d}\mu(\alpha)|\alpha\rangle\langle\bar{\alpha}|=1$ and $\int \mathrm{d}\mu(z)|z\rangle\langle \bar{z}|=1$ with $\mathrm{d}\mu(\alpha)=\mathrm{e}^{-\bar{\alpha}\alpha}\mathrm{d}^2\alpha$, $\mathrm{d}\mu(z)=\mathrm{e}^{-{\bar z}z}\mathrm{d}^2z$, and $\bar{\alpha}$ and $\bar{z}$ being the complex conjugate of $\alpha$ and $z$, we obtain
\begin{widetext}
\begin{eqnarray}
\langle \bar{\alpha}_{f},\bar{\bf z}_{f}|\rho_{\text{tot}}^{(l)}\left(t\right) |{\alpha }_{f}^{\prime },\mathbf{z}_{f}\rangle =\int\mathrm{d}\mu (\mathbf{z}_{i})\mathrm{d}\mu ({\alpha }_{i})\mathrm{d}\mu (\mathbf{z}_{i}^{\prime })\mathrm{d}\mu ({\alpha }_{i}^{\prime }) \langle \bar{\alpha}_{f},\bar{\mathbf z}_{f}|\hat{U}_l(t)|{\alpha }_{i},\mathbf{z}_{i}\rangle\langle \bar{%
\alpha}_{i},\bar{\mathbf z}_{i}|\rho_{\text{tot}}^{(l)}(0)|{\alpha }_{i}^{\prime },\mathbf{z}_{i}^{\prime }\rangle\langle \bar{\alpha}_{i}^{\prime },\bar{\mathbf z}_{i}^{\prime}|\hat{U}_l^\dag(t)|{\alpha }_{f}^{\prime },\mathbf{z}_{f}\rangle.~~~~~~
\end{eqnarray}
The forward and backward propagators can be represented in the path-integral forms as
\begin{eqnarray}
\langle \bar{\alpha}_{f},\bar{\mathbf z}_{f}|\hat{U}_l(t)|{\alpha }_{i},\mathbf{z}_{i}\rangle&=& \int D^2\alpha D^2{\bf z}\exp\Big[{\mathrm{i}\over\hbar}\Big(S_S[\alpha,\bar{\alpha}]+S_E[{\bf z},\bar{\bf {z}}]+S_I[\alpha,\bar{\alpha};{\bf z},\bar{\bf{z}}]\Big)\Big],\\
\langle \bar{\alpha}_{i}^{\prime },\bar{\mathbf z}_{i}^{\prime }|\hat{U}_l^\dag(t)|{\alpha }_{f}^{\prime },\mathbf{z}_{f}\rangle&=&\int D^2\alpha' D^2{\bf z}'\exp\Big[{-\mathrm{i}\over\hbar}\Big(S^*_S[\alpha',\bar{\alpha}']+S^*_E[{\bf z}',\bar{\bf {z}}']+S^*_I[\alpha',\bar{\alpha}';{\bf z}',\bar{\bf{z}}']\Big)\Big].
\end{eqnarray}
The initial state $\rho_\text{tot}^{(l)}(0)=\rho_l(0)\otimes\rho_\mathrm{e}^{(l)}(0)$ in the coherent-state representation is
\begin{eqnarray}
\langle \bar{\alpha}_{i},\boldsymbol{\bar{z}}_{i}|\rho _{\text{tot}}^{(l)}(0)|{\alpha }_{i}^{\prime },\mathbf{z}_{i}^{\prime }\rangle&=&\langle\bar{\alpha}_{i}|\rho _l(0)|{\alpha }_{i}^{\prime }\rangle \langle \bar{\bf z}_{i}|\rho^{(l)} _{E}(0)|{\bf z}_{i}^{\prime }\rangle\equiv\rho_l (\bar{\alpha}_{i},{\alpha }_{i}^{\prime };0)\rho_E (\bar{\bf z}_{i},{\bf z }_{i}^{\prime };0).
\end{eqnarray}
After tracing the degree of freedom of the HML, we have
\begin{eqnarray}
\rho _l(\bar{\alpha}_{f},{\alpha }_{f}^{\prime};t)&\equiv&\int \mathrm{d}\mu({\bf z}_f)\langle \bar{\alpha}_{f},\bar{\bf z}_{f}|\rho_{\text{tot}}^{(l)}\left(t\right) |{\alpha }_{f}^{\prime },\mathbf{z}_{f}\rangle=\int \mathrm{d}\mu ({\alpha }_{i})\mathrm{d}\mu ({\alpha }_{i}^{\prime })\mathcal{K}(\bar{\alpha}_{f},{\alpha }_{f}^{\prime };t|\alpha_{i},\bar{\alpha }_{i}^{\prime };0)\rho_l (\bar{\alpha}_{i},{\alpha }_{i}^{\prime };0) ,\label{fdds}
\end{eqnarray}
where the effective propagator reads
\begin{eqnarray}
\mathcal{K}(\bar{\alpha}_{f},{\alpha }_{f}^{\prime };t|\alpha_{i},\bar{\alpha }_{i}^{\prime };0)=\int D^{2}{\alpha }D^{2}{\alpha }^{\prime }\exp \{\frac{\mathrm{i}}{\hbar }(S_{S}[\bar{\alpha},{\alpha }] -S_{S}^{\ast }[\bar{\alpha}^{\prime },{\alpha }^{\prime }])\}\mathcal{F[}\bar{\alpha},{\alpha ,\bar{\alpha}}^{\prime },{\alpha }^{\prime }],\label{effpth}
\end{eqnarray}
under the boundary condition $\bar{\alpha}(t)=\bar{\alpha}_f$, $\alpha(0)=\alpha_i$, $\bar{\alpha}'(0)=\bar{\alpha}_i'$, and $\alpha'(t)=\alpha_f'$. The influence functional $\mathcal{F[}\bar{\alpha},{\alpha ,\bar{\alpha}}^{\prime },{\alpha }^{\prime }]$ collecting all the effects of the HML on the magnon is
\begin{eqnarray}
\mathcal{F[}\bar{\alpha},{\alpha ,\bar{\alpha}}^{\prime },{\alpha }^{\prime }]&=&\int \mathrm{d}\mu({\bf z}_f)\mathrm{d}\mu({\bf z }_{i})\mathrm{d}\mu({\bf z }_{i}')\int D^2{\bf z}D^2{\bf z}'\exp \{\frac{\mathrm{i}}{\hbar }(S_{E}[\bar{\bf z},{\bf z }] -S_{E}^{\ast }[\bar{\bf z}^{\prime },{\bf z }^{\prime }]\nonumber\\
&&+S_I[\alpha,\bar{\alpha};{\bf z},\bar{\bf{z}}]-S^*_I[\alpha',\bar{\alpha}';{\bf z}',\bar{\bf{z}}'])\} \rho_E (\bar{\bf z}_{i},{\bf z }_{i}^{\prime };0),\label{influc}
\end{eqnarray}
under the boundary condition $\bar{\bf z}(t)=\bar{\bf z}_f$, ${\bf z }(0)={\bf z}_i$, $\bar{\bf z}'(0)=\bar{\bf z}_i'$, and ${\bf z}'(t)={\bf z}_f'$. Performing the path integral of the variables of the HML and making the Gaussian integration, we may obtain the influence functional. Substituting the obtained influence functional into Eq. \eqref{effpth} and performing the path integral of the variables of the magnon, we obtain the effective propagator. Then the dynamics of the magnon is governed by Eq. \eqref{fdds}. 
\subsubsection*{Our system}
According to the path integral in the coherent-state representation for the harmonic oscillator system, we can exactly calculate 
\begin{eqnarray}
{\mathrm{i}\over\hbar}S_S[\alpha,\bar{\alpha}]=\frac{\bar{\alpha}(t)\alpha (t)+\bar{\alpha}(0)\alpha (0)}{2}+\int_{0}^{t}\mathrm{d}\tau \lbrack \frac{\dot{\bar{\alpha}}(\tau)\alpha(\tau) -\bar{\alpha}(\tau)\dot{\alpha}(\tau)}{2}-\frac{\mathrm{i}}{\hbar }H_S\left(\bar{\alpha}(\tau) ,\alpha(\tau) \right) ],\label{hsss1}\\
{-\mathrm{i}\over\hbar}S^*_S[\alpha',\bar{\alpha}']=\frac{\bar{\alpha}^{\prime }(t)\alpha^{\prime }(t)+\bar{\alpha}^{\prime }(0)\alpha ^{\prime }(0)}{2} +\int_{0}^{t}\mathrm{d}\tau \frac{\bar{\alpha}^{\prime }(\tau)\dot{\alpha}^{\prime }(\tau)-\dot{\bar{\alpha}}^{\prime }(\tau)\alpha ^{\prime }(\tau)}{2}+\frac{\mathrm{i}}{\hbar }H_S\left(\bar{\alpha}^{\prime }(\tau),\alpha ^{\prime}(\tau)\right) ],\label{hsss2}\\
{\mathrm{i}\over\hbar}S_E[{\bf z},\bar{\bf z}]=\sum_k\frac{\bar{z}_k(t)z_k (t)+\bar{z}_k(0)z_k (0)}{2}+\int_{0}^{t}\mathrm{d}\tau \lbrack \frac{\dot{\bar{z}}_k(\tau)z_k(\tau) -\bar{z}_k(\tau)\dot{z}_k(\tau)}{2}-\frac{\mathrm{i}}{\hbar }H_E\left(\bar{z}_k(\tau) ,z_k(\tau) \right) ],\label{hxx1}\\
{-\mathrm{i}\over\hbar}S^*_E[{\bf z}',\bar{\bf z}']=\sum_k\frac{\bar{z}_k^{\prime }(t)z_k^{\prime }(t)+\bar{z}_k^{\prime }(0)z_k ^{\prime }(0)}{2} +\int_{0}^{t}\mathrm{d}\tau \frac{\bar{z}_k^{\prime }(\tau)\dot{z}_k^{\prime }(\tau)-\dot{\bar{z}}_k'(\tau)z_k ^{\prime }(\tau)}{2}+\frac{\mathrm{i}}{\hbar }H_E\left(\bar{z}_k^{\prime }(\tau),z_k ^{\prime}(\tau)\right) ],\\
{\mathrm{i}\over\hbar}S_I[\alpha,\bar{\alpha};{\bf z},\bar{\bf z}]=\sum_k\int_{0}^{t}\mathrm{d}\tau \lbrack-\frac{\mathrm{i}}{\hbar }H_I\left(\bar{z}_k(\tau) ,\bar{\alpha}(\tau),z_k(\tau),\alpha(\tau) \right) ],\\
{-\mathrm{i}\over\hbar}S_I[\alpha',\bar{\alpha}';{\bf z}',\bar{\bf z}']=\sum_k\int_{0}^{t}\mathrm{d}\tau \lbrack\frac{\mathrm{i}}{\hbar }H_I\left(\bar{z}_k(\tau)' ,\bar{\alpha}'(\tau),z_k'(\tau),\alpha'(\tau) \right) ],\label{hxx2}
\end{eqnarray}
where $H_S(\bar{\alpha},\alpha)=\hbar[\omega_0+A(t)]\bar{\alpha}\alpha$, $H_E(\bar{z}_k,z_k)=\hbar\omega_k\bar{z}_kz_k$, and $H_I(\bar{z}_k,\bar{\alpha},z_k,\alpha)=-\hbar g_k(\bar{z}_k\alpha+z_k\bar{\alpha})$. It is found that, for our harmonic-oscillator system interacting with the bosonic environment, only the quadratic forms of the variables are involved. Therefore, the path integral and the Gaussian integration to the variables of the HML in Eq. \eqref{influc} can be exactly evaluated by the saddle-point method as
\begin{eqnarray}
&&\mathcal{F}[\bar{\alpha} ,\alpha ,\bar{\alpha} ^{\prime },\alpha ^{\prime }]= \exp\Big\{\int_{0}^{t}\mathrm{d}\tau  \int_{0}^\tau\mathrm{d}\tau ^{\prime }\big\{ F(\tau-\tau')[\bar{\alpha}'(\tau)-\bar{\alpha}(\tau)]\alpha(\tau')+F^*(\tau-\tau')\bar{\alpha}'(\tau')[\alpha(\tau)-\alpha'(\tau)]\big\}\Big\}.~~~ \label{inff}
\end{eqnarray}%
where $F(s)=\sum_k|g_k|^2\mathrm{e}^{-\mathrm{i}\omega_ks}=\int J(\omega)\mathrm{e}^{-\mathrm{i}\omega s}d\omega$ and the initial condition $\rho_E(\bar{\bf z}_i,{\bf z}_i';0)=\langle\bar{\bf z}_i|\{0_k\}_l\rangle\langle \{0_k\}_l|{\bf z}_i'\rangle=1$ has been used.

\begin{proof}
Substituting the coherent-state representation $\rho_E(\bar{\bf z}_i,{\bf z}_i';0)=\langle\bar{\bf z}_i|\{0_k\}_l\rangle\langle \{0_k\}_l|{\bf z}_i'\rangle=1$ of the initial state of $\rho^{(l)}_E(0)=|\{0_k\}_l\rangle\langle\{0_k\}_l|$ of the $l$th HML and Eqs. \eqref{hxx1}-\eqref{hxx2} into Eq. \eqref{influc}, we obtain
\begin{eqnarray}
&& \mathcal{F}[\bar{\alpha} ,\alpha ,\bar{\alpha} ^{\prime },\alpha ^{\prime }]=\int \mathrm{d}\mu({\bf z}_f)\mathrm{d}\mu({\bf z }_{i})\mathrm{d}\mu({\bf z }_{i}') \prod_k\mathcal{A}_k[\bar{z}_k,z_k;\bar{z}_k',{z}_k'],\label{dac1}\\
&& \mathcal{A}_k[\bar{z}_k,z_k;\bar{z}_k',{z}_k']=\int D^2z_kD^2z_k'\exp\Big\{{\bar{z}_k(t)z_k (t)+\bar{z}_k(0)z_k (0)+\bar{z}'_k(t)z'_k (t)+\bar{z}'_k(0)z'_k (0)\over 2}+\int_0^t\mathrm{d}\tau\big\{{\dot{\bar{z}}_k(\tau)z_k(\tau) \over 2}\nonumber\\
&&~~-{\bar{z}_k(\tau)\dot{z}_k(\tau)\over 2}+{\bar{z}'_k(\tau)\dot{z}'_k(\tau) -\dot{\bar{z}}'_k(\tau)z'_k(\tau)\over 2}-\mathrm{i}\omega_k[{\bar z}_k(\tau)z_k(\tau)-{\bar z}'_k(\tau)z'_k(\tau)]+ig_k[\bar{z}_k(\tau)\alpha(\tau)-\bar{z}_k'(\tau)\alpha(\tau)+\text{c.c.}]\big\}\Big\}.~~~~~~~~\label{act1}
\end{eqnarray}
Expanding the paths near the saddle-point paths as $z_k(\tau)=z_k^\text{s}(\tau)+\delta z_k(\tau)$, $z'_k(\tau)=z_k^{\prime\text{s}}(\tau)+\delta z'_k(\tau)$, and their respective conjugates, we have $\mathcal{A}_k[\bar{z}_k,z_k;\bar{z}_k',{z}_k']=\mathcal{A}_k[\bar{z}^\text{s}_k,z^\text{s}_k;\bar{z}^{'\text{s}}_k,{z}^{'\text{s}}_k]+\delta \mathcal{A}_k+\delta^2 \mathcal{A}_k$. Here $\mathcal{A}_k[\bar{z}^\text{s}_k,z^\text{s}_k;\bar{z}^{'\text{s}}_k,{z}^{'\text{s}}_k]$ is the action contributed by the saddle-point paths, $\delta \mathcal{A}_k$ and $\delta^2 \mathcal{A}_k$ are the contributions of the first- and second-order deviation terms to $\mathcal{A}_k[\bar{z}^\text{s}_k,z^\text{s}_k;\bar{z}^{\prime\text{s}}_k,{z}^{\prime\text{s}}_k]$. It is interesting to find that $\delta^2 \mathcal{A}_k$ does not depend on the saddle-point paths. It means that $\delta^2 \mathcal{A}_k$ only contributes a constant to $\mathcal{A}_k[\bar{z}_k,z_k;\bar{z}_k',{z}_k']$, which can be collected into the normalization constant. The saddle-point paths satisfy  $\delta \mathcal{A}_k=0$, which results in\begin{eqnarray}
\mathrm{i}\dot{z}^\text{s}_k(\tau) &=&\omega_k z^\text{s}_k(\tau)-g_k\alpha(\tau) \text{, \ \ \ \ \ }-\mathrm{i}\dot{\bar{z}}^\text{s}_k(\tau)=\omega _k\bar{z}^\text{s}_k(\tau)-g_k \bar{\alpha}(\tau); \label{ddf1} \\
\mathrm{i}\dot{z}_k^{\prime\text{s} }(\tau) &=&\omega_k z^{\prime\text{s} }_k(\tau)-g_k^{\ast }\alpha ^{\prime }(\tau)
\text{, \ \ \ \ \ }-\mathrm{i}\dot{\bar{z}}_k^{\prime\text{s} }(\tau)=\omega _k\bar{z}_k^{\prime\text{s} }(\tau)-g_k \bar{\alpha}^{\prime }(\tau);  \label{ddf2}
\end{eqnarray}%
with $z^{\text{s} }_k\left( 0\right) =z_{k,i}$, $\bar{z}^{\text{s} }_k\left( t\right) =%
\bar{z}_{k,f}$, $\bar{z}_k^{\prime\text{s} }\left( 0\right) =\bar{z}_{k,i}$, $%
z^{\prime\text{s} }_k\left( t\right) =z_{k,f}$. Their solutions are
\begin{eqnarray}
z^\text{s}_k\left( \tau \right) &=&z_{k,i}\mathrm{e}^{-\mathrm{i}\omega_k \tau }+\mathrm{i}g_k\mathrm{e}^{-\mathrm{i}\omega_k
\tau }\int_{0}^{\tau }\mathrm{d}t^{\prime }\mathrm{e}^{\mathrm{i}\omega_k t^{\prime }}\alpha \left(
t^{\prime }\right) ,\label{sdp1} ~~~~
\bar{z}^\text{s}_k\left( \tau \right)=\bar{z}_{k,f}\mathrm{e}^{-\mathrm{i}\omega_k (t-\tau )}+\mathrm{i}g_k
\mathrm{e}^{\mathrm{i}\omega_k \tau }\int_{\tau }^{t}\mathrm{d}t^{\prime }\mathrm{e}^{-\mathrm{i}\omega_k t^{\prime }}\bar{%
\alpha}\left( t^{\prime }\right) , \\
z_k^{\prime\text{s} }\left( \tau \right) &=&z_{k,f}\mathrm{e}^{\mathrm{i}\omega _k(t-\tau )}-\mathrm{i}g_k \mathrm{e}^{-\mathrm{i}\omega_k \tau }\int_{\tau }^{t}\mathrm{d}t^{\prime }\mathrm{e}^{\mathrm{i}\omega_k t^{\prime }}\alpha
^{\prime }\left( t^{\prime }\right) , ~~
\bar{z}_k^{\prime\text{s} }\left( \tau \right) =\bar{z}_{k,i}^{\prime }\mathrm{e}^{\mathrm{i}\omega_k \tau
}-\mathrm{i}g_k \mathrm{e}^{\mathrm{i}\omega_k \tau }\int_{0}^{\tau }\mathrm{d}t^{\prime }\mathrm{e}^{-\mathrm{i}\omega_k
t^{\prime }}\bar{\alpha}^{\prime }\left( t^{\prime }\right) .\label{sdp2}
\end{eqnarray}
The substitution of Eqs. \eqref{sdp1}-\eqref{sdp2} into $\mathcal{A}_k[\bar{z}^\text{s}_k,z^\text{s}_k;\bar{z}^{\prime\text{s}}_k,{z}^{\prime\text{s}}_k]$ results in
\begin{eqnarray}
&&\mathcal{A}_k[\bar{z}_k,z_k;\bar{z}^{\prime}_k,{z}^{\prime}_k]=\exp \Big\{\bar{z}_{k,f}z_{k,i}\mathrm{e}^{-\mathrm{i}\omega_k t}+z_{k,f}\bar{z}_{k,i}^{\prime }\mathrm{e}^{\mathrm{i}\omega_k t}+\mathrm{i}g_k\int_{0}^{t}\mathrm{d}\tau\big[ \mathrm{e}^{-\mathrm{i}\omega_k (t-\tau )}\bar{z}_{k,f}\alpha \left( \tau \right) + \mathrm{e}^{-\mathrm{i}\omega_k \tau }z_{k,i}\bar{\alpha}\left( \tau \right)\nonumber \\
&& ~~~~- \mathrm{e}^{\mathrm{i}\omega_k(t-\tau )}z_{k,f}\bar{\alpha}^{\prime }\left( \tau \right)- \mathrm{e}^{\mathrm{i}\omega_k\tau }\bar{z}_{k,i}^{\prime }\alpha ^{\prime }\left( \tau \right) \big]  -g_k^{2}\int_{0}^{t}\mathrm{d}\tau \int_{0}^{\tau}\mathrm{d}\tau ^{\prime }\big[\mathrm{e}^{\mathrm{i}\omega_k (\tau -\tau^{\prime })}\bar{\alpha}^{\prime }\left(\tau ^{\prime }\right) \alpha ^{\prime }\left( \tau \right)+\mathrm{e}^{-\mathrm{i}\omega_k (\tau -\tau ^{\prime })}\bar{\alpha}\left( \tau \right) \alpha \left( \tau ^{\prime }\right)\big]\Big\},~~~~~~\label{act4}
\end{eqnarray}where the constant contributed by $\delta^2 \mathcal{A}_k$ has been absorbed by the normalization determined in the final step. Substituting Eq. \eqref{act4} into \eqref{dac1} and performing the Gaussian integration, we can obtain Eq. \eqref{inff} straightforwardly. 
\end{proof}

Substituting Eq. \eqref{inff} into Eq. \eqref{effpth}, we obtain
\begin{equation}
     \mathcal{K}(\bar{\alpha}_{f},{\alpha }_{f}^{\prime };t|%
\alpha_{i},\bar{\alpha }_{i}^{\prime };0)=\exp\{c(t)\bar{\alpha}_f\alpha_i+c^*(t)\bar{\alpha}_i'\alpha_f'+[1-|c(t)|^2]\bar{\alpha}_i'\alpha_i\},
\end{equation}
where $c(t)$ satisfies the same equation as Eq. \eqref{cctd}. 

\begin{proof}
   Substituting Eqs. \eqref{inff}, \eqref{hsss1}, and \eqref{hsss2} into Eq. \eqref{effpth}, we have
   \begin{eqnarray}
&&\mathcal{K}(\bar{\alpha}_{f},\alpha _{f}^{\prime };t|\bar{\alpha}%
_{i}^{\prime },\alpha _{i};0) =\int D^{2}\alpha D^{2}\alpha ^{\prime }\exp \{\frac{\bar{\alpha}\left(
t\right) \alpha \left( t\right) +\bar{\alpha}\left( 0\right) \alpha \left(
0\right) +\bar{\alpha}^{\prime }(t)\alpha ^{\prime }\left( t\right) +\bar{%
\alpha}^{\prime }(0)\alpha ^{\prime }\left( 0\right) }{2}-\tilde{\mathcal A}[\bar{\alpha},\alpha;\bar{\alpha}',\alpha']\},\label{dfd12}\\
&&\tilde{\mathcal A}[\bar{\alpha},\alpha;\bar{\alpha}',\alpha']=\int_0^t\mathrm{d}\tau\Big\{
\frac{\bar{\alpha}(\tau)\dot{\alpha}(\tau)-\dot{\bar{\alpha}}(\tau)\alpha(\tau) }{2}+%
\frac{\dot{\bar{\alpha}}^{\prime }(\tau)\alpha ^{\prime }(\tau)-\bar{\alpha}^{\prime }(\tau)%
\dot{\alpha}^{\prime }(\tau)}{2}+\mathrm{i}[\omega _{0}+A(\tau)][\bar{\alpha}(\tau)\alpha (\tau)-\bar{\alpha}%
^{\prime }(\tau)\alpha ^{\prime }(\tau)]\nonumber\\&&~~~~~~~~~-\int_{0}^\tau\mathrm{d}\tau ^{\prime }\big\{ F(\tau-\tau')[\bar{\alpha}'(\tau)-\bar{\alpha}(\tau)]\alpha(\tau')+F^*(\tau-\tau')\bar{\alpha}'(\tau')[\alpha(\tau)-\alpha'(\tau)]\big\}
\Big\}.\label{ttaa}
\end{eqnarray}
Equation \eqref{dfd12} also has quadratic forms of its variables, which means that the path integral can be exactly evaluated. Expanding again each path near the saddle-point path as $\alpha(\tau)=\alpha_\text{s}(\tau)+\delta\alpha$, $\alpha'(\tau)=\alpha'_\text{s}(\tau)+\delta\alpha'$, and their respective conjugates, we have $\tilde{\mathcal A}[\bar{\alpha},\alpha;\bar{\alpha}',\alpha']=\tilde{\mathcal A}[\bar{\alpha}_\text{s},\alpha_\text{s};\bar{\alpha}'_\text{s},\alpha'_\text{s}]+\delta\tilde{\mathcal A}+\delta^2\tilde{\mathcal A}$. $\delta^2\tilde{\mathcal A}$ does not depend on the saddle-point paths and thus only contributes a normalization constant. The saddle-point paths satisfy $\delta\tilde{\mathcal A}=0$, which results in
\begin{eqnarray}
0&=&\dot{\alpha}_\text{s}(\tau)+\mathrm{i}[\omega _{0}+A(\tau)]\alpha_\text{s}(\tau) +\int_{0}^{\tau }\mathrm{d}\tau ^{\prime }F(\tau -\tau ^{\prime })\alpha_\text{s}(\tau
^{\prime }) ,\label{ap} \\
0&=&\dot{\bar{\alpha}}^{\prime }_\text{s}(\tau)-\mathrm{i}[\omega _{0}+A(\tau)]\bar{\alpha}_\text{s}^{\prime
}(\tau)+\int_{0}^{\tau}\mathrm{d}\tau ^{\prime }F^{\ast }(\tau -\tau ^{\prime })\bar{\alpha}_\text{s}^{\prime }(\tau
^{\prime }) , \label{apb}\\
0&=&-\dot{\bar{\alpha}}_\text{s}(\tau)+\mathrm{i}[\omega _{0}+A(\tau)]\bar{\alpha}_\text{s}(\tau)-\int_{0}^{t}\mathrm{d}\tau ^{\prime }F^{\ast }(\tau -\tau
^{\prime })\bar{\alpha}_\text{s}^{\prime }(\tau ^{\prime })+\int_{\tau }^{t}\mathrm{d}\tau
^{\prime }F^{\ast }(\tau -\tau ^{\prime })\bar{\alpha}_\text{s}(\tau ^{\prime }) ,\label{appc}
\\
0&=&-\dot{\alpha}^{\prime }_\text{s}(\tau)-\mathrm{i}[\omega _{0}+A(\tau)]\alpha ^{\prime }_\text{s}(\tau)-\int_{0}^{t}\mathrm{d}\tau ^{\prime }F(\tau -\tau
^{\prime })\alpha_\text{s} (\tau ^{\prime })+\int_{\tau }^{t}\mathrm{d}\tau ^{\prime }F(\tau
-\tau ^{\prime })\alpha_\text{s} ^{\prime }\left( \tau ^{\prime }\right) ,\label{apm}
\end{eqnarray}under the boundary conditions $\alpha_\text{s}(0)=\alpha_i$, $\bar{\alpha}_\text{s}'(0)=\bar{\alpha}_i$, $\bar{\alpha}_\text{s}(t)=\bar{\alpha}_f$, and $\alpha_\text{s}'(t)=\alpha_f'$.
The substitution of Eqs. \eqref{ap}-\eqref{apm} into Eq. \eqref{ttaa} leads to $\tilde{\mathcal A}[\bar{\alpha}_\text{s},\alpha_\text{s};\bar{\alpha}'_\text{s},\alpha'_\text{s}]=0$. Therefore, the path integral of Eq. \eqref{dfd12} only has the contribution of its boundary values $\alpha(t)$, $\bar{\alpha}(0)$, $\bar{\alpha}'(t)$, and $\alpha'(0)$, which can be determined from Eqs. \eqref{ap}-\eqref{apm} as follows. 

The solution of Eq. \eqref{ap} depends on its initial value $\alpha_i$ as $\alpha_\text{s}(\tau)=\alpha_ic(\tau)$ under $c(0)=1$ and
\begin{equation}
    \dot{c}(\tau )+\mathrm{i}[\omega _{0}+A(\tau)]c(\tau )+\int_{0}^{\tau }F(\tau -\tau
^{\prime })c(\tau ^{\prime })\mathrm{d}\tau ^{\prime } =0.\label{cct}
\end{equation}
Remembering the fact $\tau\in[0,t]$ and defining $\chi(\tau)=\alpha_\text{s}(\tau)-\alpha'_\text{s}(\tau)$, we derive from Eqs. \eqref{ap} and \eqref{apm} that $\chi (\tau )=c_{1}(\tau )\chi (t)$ with $c_{1}(\tau )$
satisfying $c_{1}(t)=1$ and%
\begin{equation}
\dot{c}_{1}(\tau)+\mathrm{i}[\omega _{0}+A(\tau)]c_{1}(\tau)-\int_{\tau }^{t}\mathrm{d}\tau ^{\prime }F(\tau -\tau
^{\prime })c_{1}(\tau ^{\prime })=0.  \label{u1}
\end{equation}
Equations \eqref{cct} and \eqref{u1} readily leads to $c_1(\tau)=\bar{c}(t-\tau)$. Therefore, we obtain 
\begin{equation}\alpha_\text{s}'(\tau)=\alpha_\text{s}(\tau)-\chi(\tau)=\alpha_ic(\tau)-\bar{c}(t-\tau)[\alpha_\text{s}(t)-\alpha_\text{s}'(t)]\Rightarrow \alpha_\text{s}'(0)=\alpha_i-\bar{c}(t)[\alpha_ic(t)-\alpha_f']=[1-|c(t)|^2]\alpha_i+\bar{c}(t)\alpha_f'.
\end{equation}
The solution of Eq. \eqref{apb} reads $\bar{\alpha}_\text{s}'(\tau)=\bar{\alpha}_i'\bar{c}(\tau)$. A similar procedure as above making on Eqs. \eqref{apb} and \eqref{appc} leads to $\bar{\alpha}_\text{s}(0)=[1-|c(t)|^2]\bar{\alpha}_i'+c(t)\bar{\alpha}_f$.

Finally, we obtain
\begin{eqnarray}
   \mathcal{K}(\bar{\alpha}_{f},\alpha _{f}^{\prime };t|\bar{\alpha}%
_{i}^{\prime },\alpha _{i};0)&=&N(t)\exp[\frac{\bar{\alpha}_f \alpha_\text{s} \left( t\right) +\bar{\alpha}_\text{s}\left( 0\right) \alpha_i +\bar{\alpha}_\text{s}^{\prime }(t)\alpha ^{\prime }_f +\bar{%
\alpha}^{\prime }_i\alpha_\text{s} ^{\prime }\left( 0\right) }{2}]\nonumber\\
%&=&N(t)\exp[\frac{c( t)\bar{\alpha}_f \alpha _i +\{[1-|c(t)|^2]\bar{\alpha}_i'+c(t)\bar{\alpha}_f\} \alpha_i +\bar{c}(t)\bar{\alpha}^{\prime }_f\alpha ^{\prime }_f +\bar{\alpha}^{\prime }_i\{[1-|c(t)|^2]\alpha_i+\bar{c}(t)\alpha_f'\} }{2}]\nonumber\\
&=&N(t)\exp\{c(t)\bar{\alpha}_f\alpha_i+\bar{c}(t)\bar{\alpha}_i'\alpha_f'+[1-|c(t)|^2]\bar{\alpha}_i'\alpha_i\}.
\end{eqnarray}
where $N(t)$ is the normalization constant collecting the contributions of the second-order derivation terms $\delta^2\mathcal{A}_k$ and $\delta^2\tilde{\mathcal A}$. Thus, given any $\rho_l(0)$, we have $\rho_l(t)=\int \mathrm{d}\mu(\alpha_f)\mathrm{d}\mu(\alpha_f')\rho _l(\bar{\alpha}_{f},{\alpha }_{f}^{\prime
};t)|\alpha_f\rangle\langle \bar{\alpha}_f'|$ with
\begin{equation}
\rho _l(\bar{\alpha}_{f},{\alpha }_{f}^{\prime
};t)=N(t)\int \mathrm{d}\mu ({\alpha }_{i})\mathrm{d}\mu ({\alpha }%
_{i}^{\prime })\exp\{c(t)\bar{\alpha}_f\alpha_i+c^*(t)\bar{\alpha}_i'\alpha_f'+[1-|c(t)|^2]\bar{\alpha}_i'\alpha_i\}\rho_l (\bar{\alpha}_{i},{\alpha }%
_{i}^{\prime };0).
\end{equation}
The normalization condition $\text{Tr}\rho_l(t)$ easily results in $N(t)=1$. 
\end{proof}

Based on the fact that there is no interaction between the two total systems consisting of each magnon mode and its local environment, we obtain the exact dynamics of the two magnon modes as $\rho(t)=\int \mathrm{d}\mu({\pmb \alpha}_f)\mathrm{d}\mu({\pmb \alpha}_f')\rho (\bar{\pmb \alpha}_{f},{\pmb \alpha }_{f}^{\prime
};t)|{\pmb \alpha}_f\rangle\langle \bar{\pmb\alpha}_f'|$ with 
\begin{equation}
\rho (\bar{\pmb \alpha}_{f},{\pmb \alpha }_{f}^{\prime
};t)=\int \mathrm{d}\mu ({\pmb\alpha }_{i})\mathrm{d}\mu ({\pmb\alpha }%
_{i}^{\prime })\exp\{\sum_{l=1}^{2}[c(t)\bar{\alpha}_{lf}\alpha_{li}+\bar{c}(t)\bar{\alpha}_{li}'\alpha_{lf}'+[1-|c(t)|^{2}]\bar{\alpha}_{li}^{\prime }\alpha _{li}\}\rho (\bar{\pmb \alpha}_{i},{\pmb\alpha }%
_{i}^{\prime };0).\label{ffm11}
\end{equation}It is emphasized that Eq. \eqref{ffm11} does not resort to any approximation from the unitary dynamics of the total system under the initial state $\rho_\text{tot}=\rho(0)\otimes |\{0_k\}_1\rangle\langle \{0_k\}_1|\otimes |\{0_k\}_2\rangle\langle \{0_k\}_2|$.

From Eq. \eqref{ffm11}, we can derive an exact master equation
\begin{equation}
\dot{\rho}(t)=\sum_{l=1}^{2}\{-\mathrm{i}\Omega (t)[\hat{a}^{\dagger}_{l}\hat{a}_l,\rho(t)]+\Gamma (t)[2\hat{a}_{l}\rho(t)\hat{a}_{l}^{\dag}-\hat{a}_{l}^{\dag}\hat{a}_{l}\rho(t)-\rho(t)\hat{a}_{l}^{\dag}\hat{a}_{l}]\},
\label{masterequati}
\end{equation}
where $\Omega(t)=-\text{Im}[\dot{c}(t)/c(t)]$ and $\Gamma(t)=-\text{Re}[\dot{c}(t)/c(t)]$ denote the renormalized frequency and decay rate of the quantum system. 

\begin{proof}
Making the time derivative to Eq. \eqref{ffm11}, we have
\begin{eqnarray}
   \dot{\rho} (\bar{\pmb \alpha}_{f},{\pmb \alpha }_{f}^{\prime};t)= \{\sum_{l=1}^{2}[\dot{c}(t)\bar{\alpha}_{lf}\alpha_{li}+\dot{\bar{c}}(t)\bar{\alpha}_{li}'\alpha_{lf}'-[\dot{c}(t)\bar{c}(t)+c(t)\dot{\bar{c}}(t)]\bar{\alpha}_{li}^{\prime }\alpha _{li}\}\rho (\bar{\pmb \alpha}_{f},{\pmb \alpha }_{f}^{\prime};t).\label{tffs}
\end{eqnarray}
Equation \eqref{ffm11} also results in
$\alpha_{li}\rho (\bar{\pmb \alpha}_{f},{\pmb \alpha }_{f}^{\prime};t)={1\over c(t)}{\partial \rho (\bar{\pmb \alpha}_{f},{\pmb \alpha }_{f}^{\prime};t)\over \partial{\bar\alpha}_{lf}}$ and $\bar{\alpha}'_{li}\rho (\bar{\pmb \alpha}_{f},{\pmb \alpha }_{f}^{\prime};t)={1\over\bar{c}(t)}{\partial\rho (\bar{\pmb \alpha}_{f},{\pmb \alpha }_{f}^{\prime};t)\over\partial \alpha'_{lf}}$, substituting which into Eq. \eqref{tffs}, we have
\begin{equation}
   \dot{\rho} (\bar{\pmb \alpha}_{f},{\pmb \alpha }_{f}^{\prime};t)= \{\sum_{l=1}^{2}{\dot{c}(t)\over c(t)}\bar{\alpha}_{lf}{\partial\over \partial \bar{\alpha}_{lf} }+{\dot{\bar{c}}(t)\over\bar{c}(t)}\alpha_{lf}'{\partial\over\partial \alpha_{lf}'}-[{\dot{c}(t)\over c(t)}+{\dot{\bar{c}}(t)\over\bar{c}(t)}]{\partial^2\over\partial\bar{\alpha}_{lf}\partial\alpha'_{lf}}\}\rho (\bar{\pmb \alpha}_{f},{\pmb \alpha }_{f}^{\prime};t) .\label{ddfg}
\end{equation}
It is easy to derive
\begin{eqnarray}
    \bar{\alpha}_{lf}{\partial\rho (\bar{\pmb \alpha}_{f},{\pmb \alpha }_{f}^{\prime};t)\over \partial \bar{\alpha}_{lf} } =\langle\bar{\pmb \alpha}_f|\hat{a}_l^\dag\hat{a}_l\rho(t) |{\pmb\alpha}_f'\rangle,~\alpha_{lf}'{\partial\rho (\bar{\pmb \alpha}_{f},{\pmb \alpha }_{f}^{\prime};t)\over\partial \alpha_{lf}'} =\langle\bar{\pmb \alpha}_f|\rho(t) \hat{a}_l^\dag\hat{a}_l|{\pmb\alpha}_f'\rangle,~{\partial^2\rho (\bar{\pmb \alpha}_{f},{\pmb \alpha }_{f}^{\prime};t)\over\partial\bar{\alpha}_{lf}\partial\alpha'_{lf}} =\langle\bar{\pmb \alpha}_f|\hat{a}_l\rho(t) \hat{a}_l^\dag|{\pmb\alpha}_f'\rangle.~~~~~\label{optr}
\end{eqnarray}
The substitution of Eq. \eqref{optr} into Eq. \eqref{ddfg} results in
\begin{eqnarray}
   \langle\bar{\pmb \alpha}_f|\dot{\rho}(t) |{\pmb\alpha}_f'\rangle =\{\sum_{l=1}^{2}{\dot{c}(t)\over c(t)}\langle\bar{\pmb \alpha}_f|\hat{a}_l^\dag\hat{a}_l\rho(t) |{\pmb\alpha}_f'\rangle+{\dot{\bar{c}}(t)\over\bar{c}(t)}\langle\bar{\pmb \alpha}_f|\rho(t) \hat{a}_l^\dag\hat{a}_l|{\pmb\alpha}_f'\rangle-[{\dot{c}(t)\over c(t)}+{\dot{\bar{c}}(t)\over\bar{c}(t)}]\langle\bar{\pmb \alpha}_f|\hat{a}_l\rho(t) \hat{a}_l^\dag|{\pmb\alpha}_f'\rangle\},\\
   \Rightarrow\dot{\rho}(t) =\{\sum_{l=1}^{2}{\dot{c}(t)\over c(t)}\hat{a}_l^\dag\hat{a}_l\rho(t) +{\dot{\bar{c}}(t)\over\bar{c}(t)}\rho(t) \hat{a}_l^\dag\hat{a}_l-[{\dot{c}(t)\over c(t)}+{\dot{\bar{c}}(t)\over\bar{c}(t)}]\hat{a}_l\rho(t) \hat{a}_l^\dag\},
\end{eqnarray}
which is just Eq. \eqref{masterequati}. 
\end{proof}

Although Eq. \eqref{masterequati} has the Lindblad form, it does not resort to the Born-Markovian approximation. The non-Markovian effect has been self-consistently incorporated into the integro-differential equation \eqref{cct}.
\end{widetext}

We have derived the exact master equation \eqref{masterequati} from the propagation equation \eqref{ffm11}. It means that the solution of Eq. \eqref{masterequati} is exactly the form of Eq. \eqref{ffm11}. The coherent-state representation of the initial state $|\Psi(0)\rangle=\text{exp}[r(\hat{a}_{1}\hat{a}_{2}-\hat{a}^{\dagger}_{1}\hat{a}^{\dagger}_{2})]|00\rangle$ is  $\rho(\bar{\pmb\alpha}_{i},{\pmb \alpha}_{i}';0)=\cosh^{-2} r\text{exp}[-\tanh r(\bar{\alpha}_{1i}\bar{\alpha}_{2i}+\alpha_{1i}^{\prime}\alpha_{2i}^{\prime})]$, substituting which into Eq. \eqref{ffm11} and performing the Gaussian integration, we can obtain the exact time-dependent solution Eq. (19) of the periodically driven magnon system in the main text. 

\section*{Supplementary Note 2: The proof of Equation (10)}

In the case of NV center, the time-dependent state \eqref{apptmp2} evolved from the initial state $|\Psi_2(0)\rangle=|e_l,\{0_k\}_l\rangle$ governed by the Hamiltonian $\hat{H}_l(t)$ can also be written as
\begin{eqnarray}
    |\Psi_2(t)\rangle&=&\hat{\mathcal U}_l(t)|\Psi_2(0)\rangle,\label{flsct}
\end{eqnarray}
where $\hat{\mathcal U}_l(t)=\mathcal{T}\exp[-{\mathrm{i}\over\hbar}\int_0^t\hat{H}_l(\tau)d\tau]$.  Combining Eqs. \eqref{apptmp2} and \eqref{flsct}, we obtain
\begin{equation}
   c(t)=\langle e_l,\{0_k\}_l|\hat{\mathcal U}_l(t)|e_l,\{0_k\}_l\rangle. 
\end{equation}
Using $\hat{\mathcal{U}}_l(t)=\sum_{n}\mathrm{e}^{-\mathrm{i}\epsilon_n t/\hbar}|u_n(t)\rangle\langle u_n(0)|$ and $|u_n(t)\rangle=|u_n(t+T)\rangle$, we have
\begin{equation}
   c(t)=\sum_{n}\mathrm{e}^{-\mathrm{i}\epsilon_nt/\hbar}|\langle u_n(0)|e_l,\{0_k\}_l\rangle|^2
\end{equation}in the stroboscopic dynamics at $t =mT$, with $m\in\mathbb{Z}$. It clearly demonstrates that the dynamics of the periodically driven NV center characterized by $c(t)$ is intrinsically determined by the feature of the quasi-energy spectrum of the total system formed by the driven NV center and its local HML in the single-excitation subspace. 

In the case of magnon, the time evolved state of the single-excitation initial state $|\Phi(0)\rangle=|1_l,\{0_{k}\}_l\rangle$ is 
\begin{eqnarray}
|\Phi(t)\rangle&=&\hat{\mathcal{U}}_l(t)|\Phi(0)\rangle=a(t)|1_l,\{0_{k}\}_l\rangle\nonumber\\
&&+\sum_{k}f_{k}(t)\hat{\tilde f}_{k}^{l\dagger}|\varnothing_l,\{0_{k}\}_l\rangle,
\end{eqnarray}
where $a(t)\equiv\langle 1_l,\{0_{k}\}_l|\hat{\mathcal{U}}_l(t)|1_l,\{0_{k}\}_l\rangle$ is the probability amplitude of single-excitation state of the magnon. Substituting $|\Phi(t)\rangle$ into the time-dependent Schr\"odinger equation, we obtain
\begin{eqnarray}
\mathrm{i}\dot{a}(t) & =&[\omega_{0}+A(t)]a(t)+\sum_{k}g_{k}f_{k}(t),\label{daat}\\
\mathrm{i}\dot{f}_{k}(t) & =&\omega_{k}f_{k}(t)+a(t)g_{k}^{*},\label{ffdd}
\end{eqnarray}
After substituting the formal solution of Eq. \eqref{ffdd} into Eq. \eqref{daat} and using the spectral density $J(\omega)=\sum_{k}|g_{k}|^{2}\delta(\omega-\omega_{k})$, we obtain
\begin{align}
\dot{a}(t)+\mathrm{i}[\omega_{0}+A(t)]a(t)+\int_{0}^{t}F(t-\tau)a(\tau) \text{d}\tau & =0,\label{eq:3}
\end{align}
under $a(0) =1$. Here, $F(t-\tau)=\int d\omega J(\omega)\mathrm{e}^{-\mathrm{i}\omega (t-\tau)}$ is the correlation function.
The comparison of Eq. (\ref{cct}) and Eq. (\ref{eq:3}) leads to
\begin{equation}
c(t)=a(t)=\langle1_l,\{0_{k}\}_l|\hat{\mathcal{U}}_l(t)|1_l,\{0_{k}\}_l\rangle.\label{uuu}
\end{equation}
According to Floquet theorem, this evolution operator is decomposed into $\hat{\mathcal{U}}_l(t)=\sum_{\alpha}\mathrm{e}^{-\mathrm{i}\epsilon_{\alpha}t/\hbar}|u_{\alpha}(t)\rangle\langle u_{\alpha}(0)|$, where $\epsilon_{\alpha}$ and $|u_{\alpha}(t)\rangle$ satisfying $[\hat{H}(t)-\mathrm{i}\partial_{t}]|u_{\alpha}(t)\rangle=\epsilon_{\alpha}|u_{\alpha}(t)\rangle$ are the quasi-energies and Floquet eigenstates in single-excitation subspace, respectively. It converts Eq. \eqref{uuu} into
\begin{equation}
c(t)=\sum_{\alpha}\langle 1_l,\{0_k\}|u_{\alpha}(t)\rangle\langle u_{\alpha}(0)|1_l,\{0_k\}\rangle \mathrm{e}^{-\mathrm{i}\epsilon_{\alpha}t/\hbar}.\label{ctttd}
\end{equation}
The components of the continuous band in Eq. \eqref{ctttd} tends to zero in the long-time limit due to the out-of-phase interference. Therefore, we have
\begin{equation}
\lim_{t\rightarrow\infty}c(t)=\langle 1_l,\{0_k\}|u^b(t)\rangle\langle u^b(0)|1_l,\{0_k\}\rangle \mathrm{e}^{-\mathrm{i}\epsilon^{b}t/\hbar},\label{eq:5}
\end{equation}
where $|u^b(t)\rangle$ is the Floquet bound state. In the stroboscopic dynamics at $t = mT$, with $m\in \mathbb{Z}$, $|u^b(t)\rangle=|u^b(0)\rangle$ and thus $\lim_{t\rightarrow\infty}c(t)=\mathcal{Z}\mathrm{e}^{-\mathrm{i}\epsilon^{b}t/\hbar}$ with $\mathcal{Z}=|\langle u^b(0)|1_l,\{0_k\}\rangle|^2$. 
Equation (\ref{eq:5}) shows that the long-time behavior of $c(t)$ is uniquely determined by the Floquet bound states in the single-excitation space. It is interesting to see that the dominated role played by the single-excitation quasi-energy spectrum of the total system in the reduced dynamics of the periodically driven system is still valid for the magnon system.

\bibliography{Ref}